\documentclass[jgr]{agutex}
\usepackage{lineno}
\usepackage{graphicx}
\usepackage{amsmath}

\authorrunninghead{Lamy et al.}
\titlerunninghead{SKR magneto-ionic modes, beaming pattern and polarization state}

\begin{document}

\title{Emission and propagation of Saturn kilometric radiation: magneto-ionic modes, beaming pattern and polarization state}

\authors{L. Lamy\altaffilmark{1,2}, B. Cecconi\altaffilmark{1}, P. Zarka\altaffilmark{1}, P. Canu\altaffilmark{3}, P. Schippers\altaffilmark{4}, W.~S. Kurth\altaffilmark{4}, R.~L. Mutel\altaffilmark{4}, D.~A. Gurnett\altaffilmark{4},  D. Menietti\altaffilmark{4}, P. Louarn\altaffilmark{5}}

\altaffiltext{1}{Laboratoire d'Etudes et d'Instrumentation en Astrophysique, Observatoire de Paris, CNRS, Meudon, France}
\altaffiltext{2}{Space and Atmospheric Physics, Blackett Laboratory, Imperial College London, London, UK}
\altaffiltext{3}{Laboratoire de Physique des Plasmas, Palaiseau, France}
\altaffiltext{4}{Department of Physics and Astronomy, University of Iowa, Iowa City, Iowa, USA}
\altaffiltext{5}{Centre d'Etude Spatiale des Rayonnements, Universit\'e Paul Sabatier, CNRS, Toulouse, France}

\begin{abstract}
The Cassini mission crossed the source region of the Saturn kilometric radiation (SKR) on 17 October 2008. On this occasion, the Radio and Plasma Wave Science (RPWS) experiment detected both local and distant radio sources, while plasma parameters were measured in situ by the magnetometer (MAG) and the Cassini Plasma Spectrometer (CAPS). A goniopolarimetric inversion was applied to RPWS 3-antenna electric measurements to determine the wave vector {\bf k} and the complete state of polarization of detected waves. We identify broadband extraordinary (X) as well as narrowband ordinary (O) mode SKR at low frequencies. Within the source region, SKR is emitted just above the X mode cutoff frequency in a hot plasma, with a typical electron-to-wave energy conversion efficiency of $\sim1\%$ (2\% peak). The knowledge of the {\bf k}-vector is then used to derive the locus of SKR sources in the kronian magnetosphere, that shows X and O components emanating from the same regions. We also compute the associated beaming angle at the source $\theta'$~=~({\bf k,-B}) either from (i) in situ measurements or a model of the magnetic field vector (for local to distant sources) or from (ii) polarization measurements (for local sources). Obtained results, similar for both modes, suggest quasi-perpendicular emission for local sources, whereas the beaming pattern of distant sources appears as a hollow cone with a frequency-dependent constant aperture angle: $\theta'=75^\circ\pm15^\circ$ below 300~kHz, decreasing at higher frequencies to reach $\theta'(1000$~kHz)~$=50^\circ\pm25^\circ$. Finally, we investigate quantitatively the SKR polarization state, observed to be strongly elliptical at the source, and quasi-purely circular for sources located beyond approximately 2 kronian radii. We show that conditions of weak mode coupling are achieved along the ray path, under which the magneto-ionic theory satisfactorily describes the evolution of the observed polarization. These results are analyzed comparatively with the Auroral Kilometric Radiation (AKR) at Earth.
\end{abstract}

\begin{article}

\section{Introduction}

Forty years of remote and in situ measurements of Earth-orbiting spacecraft led the auroral kilometric radiation (AKR) to become understood enough to be used as a basis for investigating other planetary auroral radio emissions. AKR is a powerful non-thermal emission with complex properties generated above the terrestrial atmosphere \citep{Benediktov_SR_65,Gurnett_JGR_74} by the Cyclotron Maser Instability (CMI) \citep{Wu_ApJ_79,Treumann_AAR_06} from unstable hot electrons accelerated at energies of a few keV \citep{Louarn_JGR_90, Ergun_ApJ_00}.

%AKR is a powerful cyclotron emission generated above the terrestrial atmosphere, in tenuous high-latitude auroral regions, at frequencies slightly below the local electron gyrofrequency $f_\mathit{ce}$, between 50 and 700~kHz, corresponding to altitudes between 3 and 0.5~R$_\mathit{E}$ (1~R$_\mathit{E}=$~1 Earth radius~=~6378~km) \citep{Benediktov_SR_65,Gurnett_JGR_74}. Radio sources are distributed along magnetic field lines connected to bright auroral arcs \citep{Huff_JGR_88}.

%AKR mostly propagates in the free-space, Right-Handed polarized (RH) extraordinary (X) mode \citep{Kaiser_GRL_78,Shawhan_GRL_82}, from its source region \citep{Roux_JGR_93}, and marginally in the Left-Handed polarized (LH) ordinary (O) mode, typically 2 orders of magnitude weaker \citep{Benson_RS_84,Mellott_GRL_84}. Some trapped Z mode emission has also been recorded \citep{Gurnett_JGR_83,Oya_JGR_83,Deferaudy_GRL_87}.

%Radiated by the Cyclotron Master Instability (CMI) \citep[and references therein]{Wu_ApJ_79,Treumann_AAR_06}, kilometric waves are amplified quasi-perpendicularly to the local magnetic field \citep{Hilgers_JGR_92} primarily by unstable trapped \citep{Louarn_JGR_90}, or shell-type \citep{Ergun_ApJ_00} electron distributions, with electron-to-wave energy conversion efficiency of the order of 1\% \citep{Gurnett_JGR_74}. Then, refraction along the path of emitted waves, at the boundaries of auroral cavities or through the plasmasphere, results in a complex oblique cone of emission, non-axisymmetric and partially filled \citep{Louarn_PSS_96a,Louarn_PSS_96b,xiao_JGR_07,Mutel_GRL_08}. 

Saturn kilometric radiation (SKR) is the kronian equivalent of AKR, knowledge of which essentially lies on remote observations of Voyager (flybys in 1980 and 1981) \citep{Kaiser_SSS_84}, Ulysses (distant observations in the 1990s) and Cassini (in orbit since mid-2004) \citep{Kurth_09}. Its spectrum extends from a few kHz to 1200~kHz, for altitudes between 5 and 0.1~R$_\mathit{S}$ (1~R$_\mathit{S}$~=~1 Saturn radius~=~60268~km). Like AKR, SKR displays remote properties consistent with the CMI mechanism \citep{Zarka_JGR_98}. It is emitted mainly in the X mode, with fainter O mode \citep{Lamy_JGR_08a,Cecconi_JGR_09}. Radio sources lie along magnetic field lines globally associated with atmospheric aurorae \citep{Lamy_JGR_09}, with correlated power radiated in radio and UV ranges, and a conversion efficiency between precipitated energy and radiated SKR around 0.5\% \citep{Kurth_Nature_05,Clarke_JGR_09}. 

The apparent beaming pattern of SKR has been estimated with two different approaches. On the one hand, direct measurements showed a frequency-dependent beaming angle, varying within the range $40^\circ-50^\circ$ ($40^\circ-80^\circ$ respectively) in the southern (northern respectively) hemisphere, and slightly decreasing with frequency \citep{Cecconi_JGR_09}. On the other hand, modeling work of the SKR visibility obtained best results with a frequency-dependent beaming angle, varying from $\sim70^\circ$ at 3~kHz to $\sim60^\circ$ at 1000~kHz for both hemispheres, computed from a CMI-unstable loss cone distribution with 20~keV electrons \citep{Lamy_JGR_08b}.

SKR was first observed to be almost purely circular polarized, from instantaneous and statistical equatorial and low latitudes observations \citep{Ortega_AA_90,Cecconi_PRE_06,Lamy_JGR_08a}, but high latitude measurements then revealed that kilometric waves become strongly elliptically polarized when observed from latitudes between $30^\circ$ and $60^\circ$ \citep{Fischer_JGR_09}. 

On 17 October (day 291) 2008, the Cassini spacecraft crossed the SKR source region for the first identified time at an unusual location near midnight Local Time (LT) at a 5~R$_\mathit{S}$ distance and latitudes between $-60^\circ$ and $-70^\circ$, bringing crucial in situ measurements required for a better understanding of the emission \citep{Kurth_PRE7_10}.

On this occasion, we investigated quantitatively the local wave and plasma properties \citep{Lamy_GRL_10} (hereafter paper I), briefly summarized below. An overall source region, characterized by intense signal close to $f_\mathit{ce}$ was detected by the Radio and Plasma Wave experiment (RPWS) between 0812 and 0912~UT. Within this interval, we identified 3 unambiguous events (dividing in 8 individual RPWS measurements) strictly below $f_\mathit{ce}$, that, according to extensive AKR studies at Earth, were considered as actual traversed sources. The footprints of magnetic field lines supporting local and distant SKR sources revealed a spiral auroral oval starting from nightside very high latitudes, indicating an enhanced auroral activity, proposed by \citet{Bunce_JGR_10} to result from a solar wind compression. Like at Earth, the auroral plasma is hot and tenuous, but unlike Earth, the SKR source region was crossed at much higher altitudes (4.1~R$_\mathit{S}$), where no auroral cavity (at the edges of which AKR propagation is affected) was detected. SKR frequencies observed below $f_\mathit{ce}$ were found compatible with CMI resonance frequencies computed with the observed 6 to 9~keV hot electrons as resonant particles and for amplification perpendicular to the magnetic field. Interestingly, these hot electrons display ring or shell-like distributions similar to those responsible for AKR generation at Earth \citep{Schippers_JGR_10}, and have been used to compute CMI growth rates high enough to account for observed SKR intensities \citep{Mutel_GRL_10}. 

This event is of particular interest, for remote and local SKR sources were measured simultaneously. In this study, we extend the analysis of paper I to investigate polarization and beaming properties of SKR waves, along their propagation from their source region. After having described Cassini's instrumentation and observational parameters (section \ref{data}), as well as useful wave properties in a magnetized plasma (section \ref{waves}), we first identify and characterize the different magneto-ionic modes of SKR (section \ref{modes}). Then, we compare the location of regions from which observed modes emanate (section \ref{loc}) and investigate their beaming pattern, derived for local to distant sources (section \ref{beaming}). Finally, we quantify the SKR polarization state at the source and its evolution with propagation (section \ref{polar}) and relate it to predictions of the magneto-ionic theory. These results are compared to other auroral planetary radio emissions, especially to the terrestrial case.

\section{Instrumentation and observational parameters}
\label{data}

Together with the three electric antennas $u$, $v$ and $w$, the RPWS experiment contains a High Frequency Receiver (HFR) which measures the wave electric power spectral density between 3.5~kHz and 16.125~MHz, including the usual spectral range of SKR waves \citep{Gurnett_SSR_04}. The HFR records a set of two auto-correlations and a complex cross-correlation of input signals sensed on a pair of antennas, hereafter simply called a 2-antenna measurement. 

%(also known as +X, -X and Z, the former convention being kept in this article to avoid redundancy with plasma notations)

During the time interval investigated in this study, the HFR was set up in the 3-antenna operating mode, consisting of two consecutive quasi-instantaneous 2-antenna measurements over the $(u,w)$ and $(v,w)$ pairs of monopoles. This simulates a real 3-antenna observation when the observed signal does not vary significantly between both 2-antenna measurements.

Under the point source assumption, a goniopolarimetric (GP) inversion (compounding the words goniometric - for angles - and polarimetric - for polarization - analysis) applied to each 3-antenna measurement \citep{Cecconi_RS_05} derives the six physical parameters of the observed wave, namely its full state of polarization, defined by the four Stokes parameters S, Q, U and V \citep{Kraus_66}, and, with the additional hypothesis of transverse electromagnetic waves ({\bf k.E}~=~0), the {\bf k}-vector direction, defined by two angular coordinates in the spacecraft frame. As the latter is determined by the direction perpendicular to the polarization ellipse, it cannot be unambiguously derived for a purely linearly polarized wave. Out of this limit, a direction yields two possible senses of {\bf k}, among which we select the one corresponding to the source the closest to the planet. Each quasi-instantaneous measurement at a given frequency thus characterizes the radio wave radiated by the most intense point source detected at that time and frequency.

Over the 3.5 to 1500~kHz spectral range of interest here, HFR measurements were acquired with logarithmically spaced frequency channels distributed within three consecutive bands from 3.5 to 325~kHz, with a spectral resolution $\delta f/f=5\%$, and linearly spaced channels within the high frequency band HF1 above 325~kHz, with a fixed resolution of 25~kHz. 

In the supplementary material of paper I, we described how radio data were specifically processed and selected along this interval to obtain reliable measurements. We remind below selection criteria applied to the degree of circular polarization V, the signal-to-noise ratio SNR, and the $z_r$ parameter quantifying the variation of the autocorrelation on the $w$ monopole between two consecutive 2-antenna measurements: 0.05~$\le\vert$V$\vert\le$~1.1 (to remove unpolarized or aberrant measurements), SNR~$\ge$~20~dB simultaneously on each antenna of the full 3-antenna measurement, $\vert$V$\vert\ge$~0.7 when 20~$\le$~SNR~$\le$~45~dB (to remove specifically weakly circularly polarized emissions with high SNR as low frequency variable background and narrowband emissions), and $\vert z_r\vert$~$\le$~0.05 (corresponding to a tolerance of signal variation of 10~\% between 2 consecutive sets of measurements). Out of Figures \ref{fig1} and \ref{fig2} that used all available data excluding radio frequency interference, all other Figures were built from the above data selection.

The RPWS experiment also includes three orthogonal search coil magnetic antennas aligned with the $x$, $y$ and $z$ axis of the spacecraft, and a Medium Frequency Receiver (MFR) \citep{Gurnett_SSR_04}, in charge of measuring the orthogonal magnetic components of electromagnetic waves. Along the present interval, the MFR provided measurements of the B$_z$ spectral density between 0.02~kHz and 12~kHz, with logarithmically spaced frequency channels and a spectral resolution of 7\%. 

Finally, plasma parameters were obtained along the spacecraft trajectory thanks to simultaneous magnetic field and electrons in situ measurements. High resolution observations of the magnetometer (MAG) \citep{Dougherty_SSR_04} provided the electron cyclotron frequency $f_\mathit{ce}$ and the local magnetic field vector {\bf B}, while electron moments were computed from measurements of the Electron Spectrometer (ELS) of the Cassini Plasma Spectrometer (CAPS) \citep{Young_SSR_04}, giving the plasma frequency $f_\mathit{pe}$ as well as the temperature and the total energy carried by hot electrons.

%Finally, plasma parameters were obtained along the spacecraft trajectory thanks to simultaneous magnetic field and electrons in situ measurements. High resolution observations of the magnetometer (MAG) \citep{Dougherty_SSR_04} were used to derive the electron cyclotron frequency $f_\mathit{ce}$ and the magnetic field direction, while the Electron Spectrometer (ELS) of the Cassini Plasma Spectrometer (CAPS) \citep{Young_SSR_04} provided the plasma frequency $f_\mathit{pe}$, the temperature of hot electrons and the total energy carried by the observed electrons.

\section{Waves in a magnetized plasma}
\label{waves}

In this section, we recall some useful wave properties in a tenuous and magnetized plasma ($f_\mathit{pe}\ll f_\mathit{ce}$), as the one observed in the kronian auroral region.

\subsection{Cold plasma dispersion and magneto-ionic theory}

It is well known that electromagnetic waves can propagate in a magnetized plasma in several characteristic (or natural) modes, defined by indices of refraction $N$ which are solutions of the dispersion equation. For a homogeneous cold plasma, neglecting the ions motion (high frequency approximation) and collisions, the dispersion relation, known as the Altar-Appleton-Hartree or Appleton-Lassen equation \citep{Appleton_32,Lassen_27}, is:

\begin{equation}
N_\mathit{_{X,O}}^2 = 1-\frac{2X(1-X)}{2(1-X)-Y^2\sin^2\theta\mp\Delta} 
\label{eq_N}
\end{equation}

\begin{equation*} 
\text{with } \Delta = \sqrt{Y^4\sin^4\theta+4Y^2(1-X)^2\cos^2\theta}
\end{equation*}
\\
\
where $X=(f_\mathit{pe}/f)^2$ and $Y=f_\mathit{ce}/f$ are the characteristic frequencies of the medium where $f$ is the wave frequency, $\theta=({\bf k},{\bf B})$ is the wave propagation angle, and the signs $-$ and $+$ refer to the extraordinary and ordinary solutions, respectively. This expression is the basis of the magneto-ionic theory.

In the general case of oblique propagation, the upper branch of the extraordinary solution and the ordinary solution are often labelled R-X and L-O modes, as an intermediate case between R,L modes (solutions for parallel propagation) and X,O ones (solutions for perpendicular propagation) \citep{Goertz_95}. As the lower branch of the extraordinary solution, namely the Z mode, also displays R-X characteristics above $f_\mathit{pe}$ (and L-X below), we will hereafter use the simple X, O and Z denomination for clarity.

Superluminous ($N\le1$) X and O modes are defined above their cutoff frequency, $f_\mathit{X}\sim f_\mathit{ce}(1+(f_\mathit{pe}/f_\mathit{ce})^2)$ and $f_\mathit{O}=f_\mathit{pe}$ respectively, and tend asymptotically toward a light wave at high frequencies ($N=1$), so that they can ultimately propagate freely through space. In contrast, the Z mode, subluminous above $f_\mathit{pe}$ and superluminous below, is trapped between its cutoff frequency $f_\mathit{Z}\sim f_\mathit{pe}^2/f_\mathit{ce}$ and the upper hybrid (resonance) frequency $f_\mathit{UH}={(f_\mathit{pe}^2+f_\mathit{ce}^2)^{1/2}}\sim f_\mathit{ce}(1+1/2(f_\mathit{pe}/f_\mathit{ce})^2)$. As $f_\mathit{UH}$ is always lower than $f_\mathit{X}$, a frequency gap separates X and Z components. 

In addition to the refractive index, an important characteristic of magneto-ionic modes given by the dispersion equation is the wave polarization, which is elliptical in the general case \citep{Melrose_80,Budden_85}. The transverse component of the polarization, defined as the axial ratio T of the polarization ellipse in the plane perpendicular to {\bf k}, is related to the refractive index by:

\begin{equation}
N_\mathit{_{X,O}}^2 = 1-\frac{XT_\mathit{_{X,O}}}{T_\mathit{_{X,O}}-Y\cos\theta}
\end{equation}\\

T can be expressed as a function of $X$, $Y$ and $\theta$:

\begin{equation}
T_\mathit{_{X,O}} = A\mp sign(A)\sqrt{1+A^2}
\label{eq_AR}
\end{equation}

\begin{equation*}
\text{with } A = \frac{Y\sin^2\theta}{2(1-X)\cos\theta}
\end{equation*}
\\
\
where the signs $-$ and $+$ refer to the extraordinary and ordinary solutions, respectively. The longitudinal component of the wave polarization (along {\bf k}) is neglected, being of the order of XY. The solutions of equation \ref{eq_AR} are orthogonal (T$_\mathit{_{X}}$T$_\mathit{_{O}}=-1$). 

The axial ratio is related to the Stokes parameters \citep{Kraus_66}, and the normalized degree of circular polarization V is given by:

\begin{equation}
V = \frac{2T}{1+T^2}
\label{eq_V}
\end{equation}

We retrieve the limiting cases of purely circularly polarized waves for parallel propagation (T~$=\pm1$, V~$=\pm1$), and purely linearly polarized ones for perpendicular propagation (T~$=0,\pm\infty$, V~=~0), whereas the wave polarization is elliptical in the general case.

The sign of V and T indicates the sense of polarization: for $A\ge0$, T$_\mathit{_X}$,V$_\mathit{_X}\le0$ corresponds to a RH polarized wave, and T$_\mathit{_O}$,V$_\mathit{_O}\ge0$ to a LH polarized wave, while the opposite situation occurs for $A\le0$. Thus, for $\theta\ge\pi/2$, which corresponds to the general case of SKR emission in the southern kronian hemisphere, X and subluminous Z mode are LH polarized while O and superluminous Z mode are RH polarized.

% thus display a sense of polarization opposite to the one of O and superluminous Z mode. Transposed to the present study, where radio emissions of the southern magnetic hemisphere are expected to propagate mainly with $\theta\ge\pi/2$, X and O modes are LH and RH polarized respectively, whereas Z mode is LH polarized above $f_\mathit{pe}$ and RH below.

\subsection{Hot plasma dispersion}
\label{waves_hot} 

Similarly to the terrestrial auroral cavities, the kronian auroral region is dominated by weakly relativistic (hot) electrons while thermal (cold) electrons are typically 5 times less dense. The cold plasma approximation is consequently very rough and relativistic effects need to be taken into account. In the frame of AKR generation, several authors solved the wave dispersion equation in a hot plasma, and showed that this essentially results in lower characteristic frequencies of the extraordinary branches, which in turn favors CMI resonance close to $f_\mathit{ce}$ \citep{Wu_PF_81,Wong_JPP_82,Pritchett_JGR_84,Lequeau_PF_84,Lequeau_JGR_84,Winglee_ApJ_85, Louarn_PSS_96b}.

For the purpose of this article, we are interested in the modified X mode cutoff, that was expressed from various typical hot electron distributions as:

\begin{equation}
f_\mathit{X} \sim f_\mathit{ce}[1+\alpha(\frac{f_\mathit{pe}}{f_\mathit{ce}})^2+\beta(\frac{v_\mathit{e}}{c})^2]
\label{eq_fx}
\end{equation}

where $v_\mathit{e}$ is the characteristic velocity of dominant hot electrons, $c$ is the speed of light and ($\alpha$,$\beta$) are coefficients related to the chosen distribution function. \citet{Winglee_ApJ_85} computed $f_\mathit{X}$ for a Maxwellian ($\alpha=1,\beta=-5/2$) and a Dory, Guest and Harris (DGH) distribution at first order ($\alpha=1,\beta=-9/2$) under the condition $(v_\mathit{e}/c)^2\le(f_\mathit{pe}/f_\mathit{ce})^2$. \citet{Louarn_PSS_96b} derived a more general expression of $f_\mathit{X}$ for an idealized ring-like distribution ($\alpha=1/2,\beta=-1/2$), and underlined that more realistic electron distributions would not affect it significantly. Equation \ref{eq_fx} shows that $f_\mathit{X}$ can be lower than $f_\mathit{ce}$ for energetic enough electrons, in agreement with AKR observations.

\begin{figure*}
\centering\includegraphics[width=\textwidth]{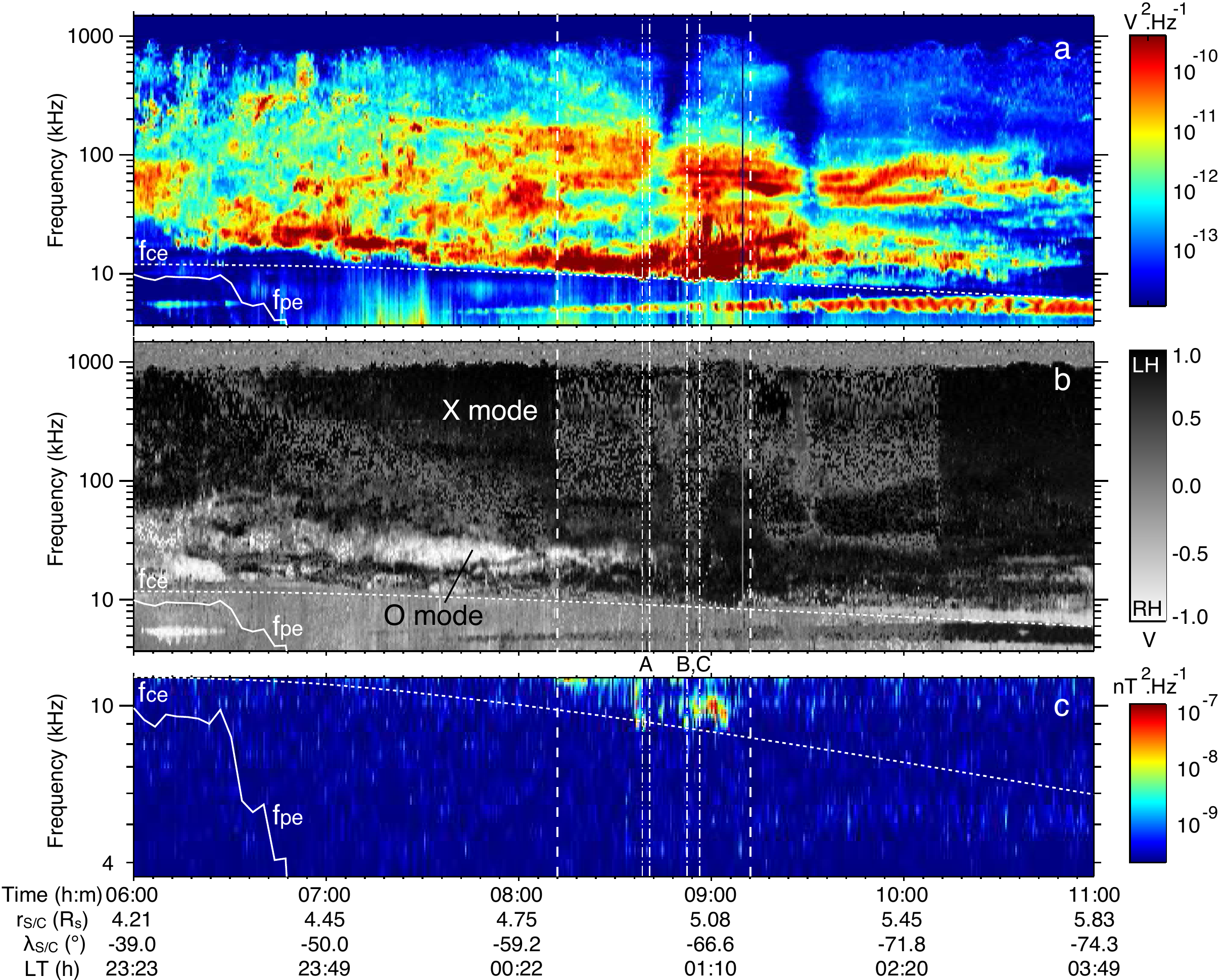}
\caption{RPWS-HFR dynamic spectra of (a) electric spectral density recorded by the $w$ monopole, (b) degree of circular polarization V, as derived from a 3-antenna GP inversion, over day 291 of year 2008 between 0600 and 1100~UT. (c) RPWS-MFR dynamic spectrum of magnetic spectral density recorded simultaneously along the z axis of the spacecraft, between 3 and 12~kHz. A daily background, computed for each frequency, has been subtracted to individual time-frequency measurements. White dashed and solid curves indicate $f_\mathit{ce}$ and $f_\mathit{pe}$, as derived from MAG and CAPS measurements. Vertical dashed lines remind the source region identified in paper I, where SKR is enhanced close to $f_\mathit{ce}$, while vertical dotted-dashed lines delimit events A, as well as B and C, for which $f<f_\mathit{ce}$.}
\label{fig1}
\end{figure*}

\section{SKR modes of emission and propagation}
\label{modes}

RPWS observations of day 291, 2008 are displayed in Figure \ref{fig1}, between 0600 and 1100~UT, in the form of three dynamic spectra, bringing complementary informations on the nature of Saturn kilometric radiation. Figure \ref{fig1}a displays the spectral power density recorded on the $w$ monopole between 3.5 and 1500~kHz. While intense electromagnetic narrowband (NB) emissions are seen around 5~kHz \citep{Ye_JGR_10}, together with weak sporadic broadband emission below 10~kHz related to auroral hiss \citep{Kopf_GRL_10}, SKR is dominant between $\sim f_\mathit{ce}$ and 1000~kHz. As identified in paper I, the overall source region corresponds to enhanced SKR close to $f_\mathit{ce}$ between 0812 and 0912~UT (vertical dashed lines), with three successive events A, B and C measured at frequencies below $f_\mathit{ce}$ (delimited by vertical dotted-dashed lines).

\subsection{Magneto-ionic modes}
\label{magnetomodes}

Similarly to AKR, remote observations identified SKR propagating in both free-space X and O modes, but unlike AKR, no Z mode SKR has been detected so far. 

Here, we can directly identify magneto-ionic modes in Figure \ref{fig1}b, that displays a dynamic spectrum of circular polarization, thanks to wave characteristics (frequency domain and sense of polarization) discussed in section \ref{waves}. LH polarized emission (black, V~$\ge0$), dominant over the whole SKR spectrum and continuously observed from local ($f\sim f_\mathit{ce}$) to distant ($f\ge f_\mathit{ce}$) sources, is consistent with predominant X mode radiated and propagating at $\theta\ge\pi/2$ (approximately outward the planet), according to equation \ref{eq_V}. At Earth, Z mode AKR corresponds to broadband weak emission \citep[and references therein]{Deferaudy_GRL_87} that can extend significantly below $f_\mathit{ce}$, which is not observed here. However, the HFR spectral resolution ($\delta f=$~450~Hz at 9~kHz) is not sufficient to resolve the forbidden frequency gap between X and Z modes ($f_\mathit{X}-f_\mathit{UH}\sim45$~Hz, using $f_\mathit{pe}=900$~Hz as derived in paper I) and, therefore, we cannot formally exclude the presence of the latter at the lower edge of the X mode emission.

SKR also displays sporadic RH emission (white, V~$\le0$) below 40~kHz. This corresponds either to O mode emitted in the same direction as the main LH component ($\theta\ge\pi/2$), or to X mode radiated in the opposite direction. Indeed, at the traversed source, $\nabla B$ lies at a few degrees relative to {\bf B}, allowing an X mode propagation window of a few degrees toward the planet. Here, we favor O mode because this emission is fainter than the predominant LH X mode, and appears at the lower edge of its envelope, as expected from $f_\mathit{O}\le f_\mathit{X}$, and in agreement with previous observations at Saturn \citep{Cecconi_JGR_09} or the Earth \citep{Panchenko_RS_08}. For comparable frequencies between 15 and 30~kHz, the X/O intensity ratio reaches $\sim 10^2$. The overall SKR pattern displays highly circularly polarized emission ($|$V$|$~$\sim1$), except around $f_\mathit{ce}$, where V decreases while the emission level remains high. This is a consequence of elliptical polarization observed for local sources, that is investigated in details in section \ref{polar}. 

Finally, we note that low frequency NB emissions and auroral hiss display clear Z mode characteristics, with LH polarization between $f_\mathit{pe}$ ($\le 2$~kHz after 0700~UT) and $f_\mathit{ce}$, and RH polarization below $f_\mathit{pe}$, indicating wave propagation at $\theta\ge\pi/2$.

%In summary, the coexistence of both X and O modes of emission dominated by X waves confirm previous results, themselves in agreement with AKR characteristics at Earth, whereas we were not

\subsection{Electromagnetic waves}

Figure \ref{fig1}c shows a dynamic spectrum of the wave magnetic field component measured along the $z$ axis of the spacecraft between 3.5 and 12~kHz. In spite of a lower sensitivity than electric measurements of Figure \ref{fig1}a, a clear signal is detected within the SKR envelope, as well as a fainter one within the trace of NB emission at the end of the interval, giving a direct evidence of the electromagnetic nature of these emissions. In addition, Figure \ref{fig1}c confirms the observation of SKR events $f<f_\mathit{ce}$ at timings consistent with the ones identified in paper I. 

Due to the uncertainty on the wave direction with respect to the $z$ axis and the lack of high SNR measurements, it was not possible to derive reliable, accurate values of the refractive index. However, the ratio of the wave electric field E and $c.$B was often close to unity at frequencies slightly above $f_\mathit{ce}$, consistent with $N\sim1$.

\begin{figure*}
\centering\includegraphics[width=0.8\textwidth]{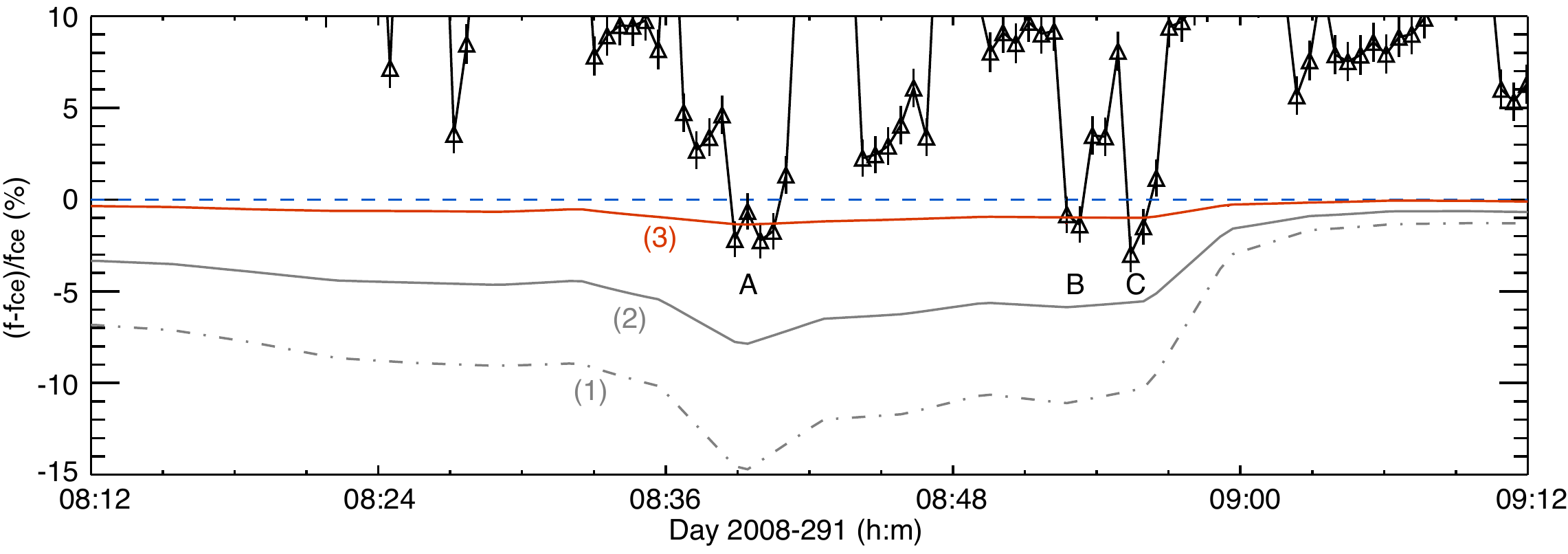}
\caption{Characteristic frequencies between 0812 and 0912~UT of day 2008-291, expressed relatively to $f_\mathit{ce}$. The SKR cutoff frequency $f_\mathit{cut}$, determined with an accuracy of 0.5~\%, is displayed in black, and $f_\mathit{ce}$, in dashed blue. The curves labelled (1), (2) and (3) plot the X mode cutoff frequency $f_\mathit{X}$ computed from (1) maxwellian, (2) DGH \citep{Winglee_ApJ_85} and (3) ring-like electron distributions \citep{Louarn_PSS_96b}, using $f_\mathit{ce}$, $f_\mathit{pe}$ and the kinetic energy of (6 to 9~keV) hot electrons, as derived from MAG observations and CAPS-ELS 200~s-averaged moments.}
\label{fig2}
\end{figure*}

\subsection{X mode cutoff}
\label{fx}

Based on remarkably similar AKR and SKR features, the CMI has been proposed as a common mechanism responsible for both radiations. Through this process, resonant electrons can amplify a background of radio waves (see for instance \citep{Zarka_JGR_86}) at frequencies close to $f_\mathit{ce}$, primarily in the X mode. To check the robustness of this interpretation, we want to verify that the SKR local emission frequency remains higher than the theoretical X mode cutoff frequency $f_\mathit{X}$.

The SKR low frequency cutoff $f_\mathit{cut}$, as determined by the lower limit of the SKR envelope, is plotted in black in Figure \ref{fig2} over the source region and relatively to $f_\mathit{ce}$. In paper I, we showed that the CMI resonance frequency matches $f_\mathit{cut}$ only for crossed sources A, B and C, distributed in 8 measurements $f_\mathit{cut}<f_\mathit{ce}$.

From in situ measurements of $f_\mathit{ce}$, $f_\mathit{pe}$ and $v_\mathit{e}$, we calculated the three types of $f_\mathit{X}$ given in section \ref{waves_hot} along the Cassini trajectory, shown in Figure \ref{fig2} by curves (1), (2) and (3). X mode cutoffs from \citet{Winglee_ApJ_85} are always lower than the CMI frequency. Nevertheless, as the condition $(v_\mathit{e}/c)^2\le(f_\mathit{pe}/f_\mathit{ce})^2$ is poorly satisfied in the source region, the real X mode cutoff should lie above curves (1) and (2). A more realistic cutoff, derived by \citet{Louarn_PSS_96b}, is given by curve (3). It matches almost all $f_\mathit{cut}$ of events A, B and C within error bars, with slightly higher average values. However, no uncertainty was considered on $f_\mathit{X}$ since it mainly depends on the technique used to derive the temperature of hot electrons. Whereas the electron moments used here were computed through a direct integration of CAPS-ELS data, averaged over 200~s to obtain a continuous series, a forward modeling of individual CAPS-ELS spectra \citep{Schippers_JGR_08} leads to less continuous but somewhat hotter temperatures (up to 1~keV more), which in turn result in $\sim$0.02\% lower $f_\mathit{X}$, in better agreement with SKR cutoff.

Investigating the regimes authorized by the CMI resonance, \citet{Louarn_PSS_96b} derived a critical electron kinetic energy $E_\mathit{cr}=(3/8)(f_\mathit{pe}/f_\mathit{ce})^2 m_\mathit{e}c^2$, above which X mode is favored over Z mode, and conversely below. In paper I, we computed ratios $f_\mathit{pe}/f_\mathit{ce}~=~0.05-0.09$ within the source region, corresponding to $E_\mathit{cr}=0.5-1.5$~keV, well below the $6-9$~keV range of dominant (and resonant) hot electrons. Finally, O mode growth rates have been shown to be significantly lower than the growth rate of dominant (X or Z) mode, whatever the $f_\mathit{pe}/f_\mathit{ce}$ ratio \citep{Wu_ApJ_79,Winglee_ApJ_85}. These results are fully consistent with the SKR modes identified above.

%Finally, the CMI process requiring low $f_\mathit{pe}/f_\mathit{ce}$, the low frequency extent of SKR spectrum, observed together with intense auroral activity and plasmoid activity in the magnetotail, may be the simple result

\begin{figure*}
\centering\includegraphics[width=0.8\textwidth]{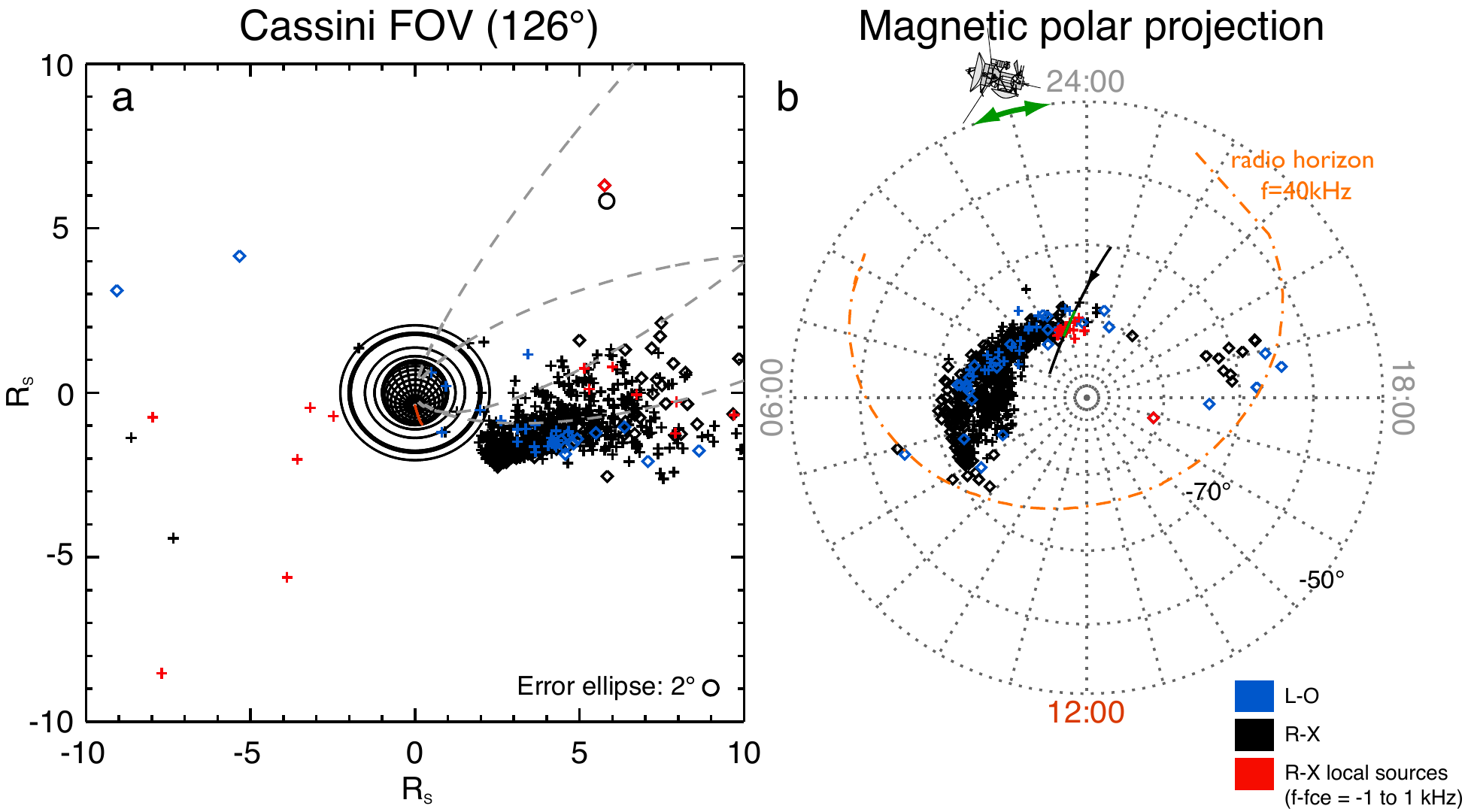}
\caption{(a) Location of radio sources as seen from Cassini, plotted with a field of view (FOV) of 126$^\circ$, and (b) magnetically projected down to the planet over the time-frequency interval [0812,0912]~UT (source region) and [$f_\mathit{ce}$-1~kHz,$f_\mathit{ce}$+40~kHz] (highest resolution). This Figure is identical to Figure 2a,b of paper I except that black and blue symbols here distinguish X (RH polarized) from O (LH polarized) mode emissions. Local X mode sources, corresponding to frequencies between $f_\mathit{ce}-1$~kHz and $f_\mathit{ce}+1$~kHz, are displayed in red. Both X and O mode SKR sources emanate from the same magnetic field lines, clustered on the dawn side, and whose footprints form a spiral shape (see also the radio map displayed by Figure 2c in paper I).}
\label{fig3}
\end{figure*}

\subsection{Electron-to-wave energy conversion efficiency}
\label{efficiency}

The energy conversion efficiency from unstable electrons to amplified radio waves has been estimated in previous studies by two methods. 

A macroscopic approach consists of comparing the total radio power radiated over its full spectrum assuming a known solid angle, to the total input (or dissipated) electron energy, assuming that the latter is a good approximation of the free energy available to the CMI resonance. Doing so, previous estimations led to typical efficiencies of 1\% for AKR \citep{Gurnett_JGR_74} and 0.5\% for SKR \citep{Kurth_Nature_05}. However, the input energy only includes precipitating electrons, while CMI-unstable electron distributions found in radio source regions involve mirrored particles, with maximum growth rates for electrons at large pitch angles, namely close to their mirror point. Thus, the input energy only gives a rough estimate of the available free energy.

Alternately, a microscopic approach directly compares radio and particle measurements within the source region using a few assumptions. At Earth, \citet{Benson_GRL_79} computed the total AKR power radiated over its full spectrum from the electric field strength measured locally and compared it to the local particle flux to derive efficiencies between 0.1 and 1\%. 

A similar computation can be performed here. In paper I, we derived a typical electric field strength of $\sim1$~mV.m$^{-1}$, estimated from the mean spectral power density measured on the single $w$ monopole and integrated over a 5~kHz wide emission. Here, we are interested in estimating the peak Poynting flux S=$\epsilon_0 cE^2/2$ and electric field amplitude close to $f_\mathit{ce}$, for the 8 individual measurements of events A, B and C, once corrected from the geometric projection of the wave electric field onto the antennas, and integrated over the instrumental bandwidth of each filter ($\sim0.5$~kHz). Therefore, we use the magnitude of the Poynting vector given by the 3-antenna GP inversion assuming $N=1$. The peak Poynting flux for the first measurement of event A reaches $9.2\times10^{-9}$~W.m$^{-2}$, corresponding to an electric field strength of 2.6~mV.m$^{-1}$, slightly higher than our previous estimation. Considering that the signal observed at 10~kHz reached the 1\% occurrence level, whose typical spectrum can be approximated by a constant signal over a 800~kHz bandwidth (see Figure 8 of \citet{Lamy_JGR_08a}), we compute the equivalent power flux radiated along an entire field line $1.5\times10^{-5}$~W.m$^{-2}$. Then, anticipating results of section \ref{beaming}, we consider that the SKR is beamed within an emission cone of solid angle $\sim0.5$~sr. The corrected power flux in $4\pi$~sr then reaches $5.9\times10^{-7}$~W.m$^{-2}$.sr$^{-1}$, or equivalently $5.9\times10^{-4}$~ergs.cm$^{-2}$.s$^{-1}$.sr$^{-1}$.

On the other hand, the available free energy can be estimated by the CAPS-ELS derived energy flux for electrons $\ge1$~keV, averaged over all anodes, assuming that hot electrons are CMI-unstable over the SKR source region. This energy flux represents more than 97\% of the total electron energy flux over the source region, which illustrates that the energy flux of cold electrons is negligible. For our example, E$_{e}(\ge1$~keV$)=0.040$~ergs.s$^{-1}$.cm$^{-2}$. Compared to the peak SKR energy radiated, we obtain a final electron-to-wave energy conversion efficiency of 1.5\%.

Table \ref{tab_efficiency} summarizes values obtained for each of the 8 measurements constituting source events A, B and C. The final mean efficiency reaches 0.9\% with a standard deviation of 0.6\%, in excellent agreement with previous estimations.

%The derived range of efficiencies includes previous estimations, but with somewhat lower average values, possibly because the electron energy was estimated over the solid angle of one single anode, and could have missed part of energetic electrons related to present events A, B and C.

\begin{table*}
\center
\begin{tabular}{c|c|c|c|c|c|c|c|c}
  SKR sources & \multicolumn{4}{c|}{A} & \multicolumn{2}{c|}{B} & \multicolumn{2}{c}{C} \\
  \hline
  DOY 2008-291 (h$:$min$:$s)&08$:$38$:$53&08$:$39$:$25&08$:$39$:$57&08$:$40$:$29&08$:$52$:$45&08$:$53$:$17&08$:$55$:$25&08$:$55$:$57\\
  \hline
  Peak Poynting flux ($\times10^{-9}$~W.m$^{-2}$)&9.2&3.5&9.3&6.0&2.3&1.2&0.96&3.6\\
  \hline
  Electric field amplitude (mV.m$^{-1}$)&2.6&1.6&2.6&2.1&1.3&0.96&0.85&1.6\\  
  \hline
  Total radiated power ($\times10^{-4}$~ergs.s$^{-1}$.cm$^{-2}$.sr$^{-1}$)&5.9&2.0&5.7&3.0&1.2&0.91&0.71&2.7\\
  \hline
    E$_{e}(\ge1$~keV) ($\times10^{-2}$~ergs.s$^{-1}$.cm$^{-2}$.sr$^{-1}$)&4.0&4.0&3.0&3.0&2.2&2.2&3.6&2.1\\  
  \hline
    Electron-to-wave efficiency (\%)&1.5&0.50&1.9&1.0&0.55&0.41&0.20&1.3\\
\end{tabular}
\caption{Table of radio and electron observables observed during events A, B and C.}
\label{tab_efficiency}
\end{table*}

\section{Source location}
\label{loc}

Under the assumption of a cyclotron emission at $f=f_\mathit{ce}$ and straight line propagation, RPWS' ability to retrieve the direction of incoming radio waves, coupled to a magnetic field model (SPV model \citep{Davis_JGR_90} including a simple current sheet \citep{Connerney_JGR_83}), enables us to derive both the spatial location of radio sources, and the beaming angle $\theta$ between the wave vector {\bf k} and the local magnetic field vector {\bf B} at the source (see \citep{Cecconi_JGR_09} for details). 

The overall spatial distribution of local and distant SKR sources, mixing all polarized components, was investigated in paper I. Here, Figure \ref{fig3} shows the location of both X (black symbols) and O (blue symbols) mode sources, as observed between 0812 and 0912~UT below $f_\mathit{ce}+30$~kHz, where both components were observed together.

Either observed from Cassini (Figure \ref{fig3}a), or projected down to the planetary atmosphere (Figure \ref{fig3}b), distant X and O emissions originate from the same range of magnetic field lines, along the visible part of a spiral auroral oval, starting from very high latitudes ($\sim-80^\circ$) close to midnight. Both components are mainly observed on the dawn side of the oval, between 0000 and 0800~LT. Nonetheless, a few X and O events also emanate from a dusk region at lower latitudes ($\sim-70^\circ$), possibly materializing the end of the spiral structure. Local sources ($f\le f_\mathit{ce}+1$~kHz, in red), identified as X mode in section \ref{modes}, map the footprint of the field lines crossed by the spacecraft on Figure \ref{fig3}b. 

\begin{figure*}
\centering\includegraphics[width=0.95\textwidth]{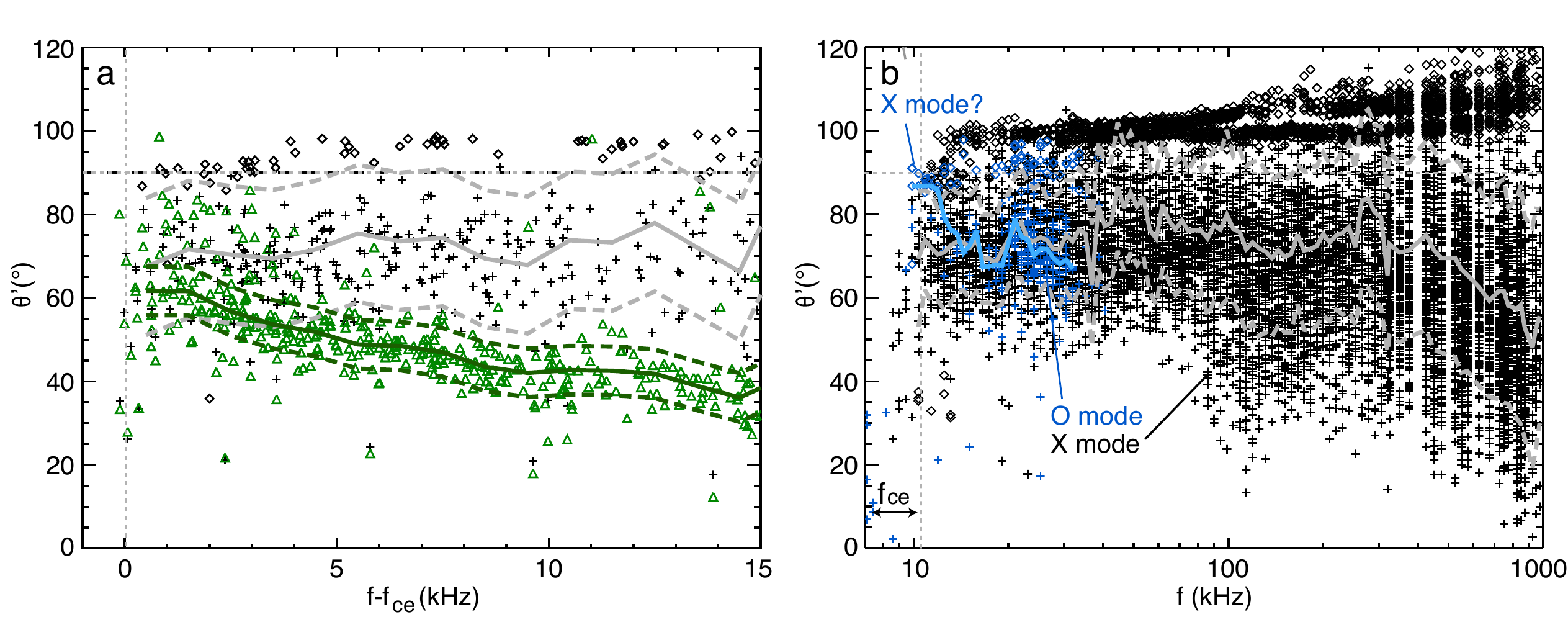}
\caption{(a) SKR beaming pattern of local sources ($f$ close to $f_\mathit{ce}$) over the interval [0812,0912]~UT, expressed in beaming angle $\theta'=$~({\bf k,-B}) between the incoming wave direction and the magnetic field vector at the source, as a function of $f-f_\mathit{ce}$. Each time-frequency {\bf k}-vector direction is derived from GP analysis, with the data selection described in section \ref{data}, whereas the magnetic field vector at the source is either given by the SPV model and a simple current sheet (black symbols), or by in situ MAG measurements (green symbols). The median value computed over all measurements (solid line), and the standard deviation (dashed lines) are plotted for each distribution. SPV-derived $\theta'$ contains two distinct distributions \citep{Cecconi_JGR_09}: reliable beaming angles are given by wave directions intercepting the surface $f=f_\mathit{ce}$ (crosses), while approximate beaming angles are estimated from wave that do not cross this surface (diamonds). Both distributions match within error bars with $\theta'\sim70^\circ$ below $f_\mathit{ce}+1$~kHz. Above this limit, MAG-derived $\theta'$ give less and less valid approximations of the beaming angle at the source and rather indicate the local wave propagation angle. (b) SKR beaming pattern of distant sources over the extended interval [0700,1100]~UT and frequencies between 7 and 1000~kHz. Only SPV-derived $\theta'$ are displayed here, separating LH (black symbols) and RH (blue symbols) polarizations, together with their median value (solid line) and standard deviation (dashed lines). In spite of scattered measurements ($\Delta\theta'\sim\pm$15$^\circ$), the median beaming angle of the X mode (solid gray) displays a clear trend, with a plateau at $\theta'(f)\sim$75$^\circ$ below 300~kHz, followed by a decrease with frequency until $\theta'$(1000~kHz)$\sim50^\circ$.}
\label{fig5}
\end{figure*}

\section{Beaming pattern}
\label{beaming}

The beaming pattern of planetary radio emissions is another crucial constraint to the generation mechanism at the source (when observed in situ), as well as to the wave propagation in the surrounding plasma (when observed remotely). In this section, we compare the SKR beaming pattern determined in situ and remotely. 

Because investigated radio sources lie in the southern magnetic hemisphere, we will hereafter rather use $\theta'=\pi-\theta=$~({\bf k,-B}) to transpose results in the usual $0^\circ$ to $90^\circ$ range, and enable comparisons with previous work.

\subsection{Local sources}

We showed in section \ref{modes} that local sources are dominantly X mode emission. In Figure \ref{fig5}a, the corresponding distribution of beaming angles $\theta'$ is derived over the interval of the source region by using both the SPV model (black) and in situ MAG observations (green), the latter being only valid locally. Results are plotted as a function of $f-f_\mathit{ce}$, corresponding to gradually increasing distances $d$ to the source.

SPV-derived results consist of two distinct distributions: reliable beaming angles are given by wave directions intercepting the surface $f=f_\mathit{ce}$ (crosses), while approximate beaming angles are estimated from wave directions that do not cross this surface (diamonds). Nonetheless, a reliable estimate of $\theta'$ is given by the median computed over all beaming angle measurements (solid gray), as detailed in appendix \ref{simu}. The difference between the black and green distributions is related to the actual angle between the modeled field vector at the source and the one measured in situ (the wave direction remaining unchanged).  

% Nevertheless, the number of available measurements below $f_\mathit{ce}$ is not sufficient to give reliable median (or average) values, while

Below $f_\mathit{ce}+1$~kHz, both techniques tend to comparable results with $\theta'=70^\circ\pm15^\circ$ for the black distribution and $\theta'=62^\circ\pm6^\circ$ for the green one (but with more and more dispersed measurements above $70^\circ$ with decreasing $f-f_\mathit{ce}$). However, below $f_\mathit{ce}$, the number of available measurements is not sufficient to give a reliable statistical estimate of $\theta'$, that displays highly dispersed values, between $30^\circ$ and $80^\circ$.

To interpret this dispersion, we remind that both determinations of $\theta'$ rely on the validity of directions given by the 3-antenna GP inversion. Indeed, the correspondence between black crosses and green triangles within a few degrees (that validates the use of the SPV model in the radio localization technique) indicates that the angular uncertainty on the modeled field is negligible compared to the scattering of {\bf k}-vectors. This scattering is likely to be the result of two main factors. First, observations close to the source region can be affected by mixed emissions from several separate intense sources. Second, as reminded in section \ref{data}, the accuracy of wave direction decreases with V, while Figure \ref{fig1}b specifically shows that V becomes low close to $f_\mathit{ce}$. These reasons can consequently limit the accuracy of the determination of $\theta'$ for the closest sources, and we will keep $\theta'=70^\circ\pm15^\circ$ as a reasonable estimate of the beaming angle observed for local sources, quasi-perpendicular to the local magnetic field vector.

Above $f_\mathit{ce}+1$~kHz, MAG-derived $\theta'(f)$ shifts toward lower values, as the local magnetic field becomes a less and less valid approximation of the magnetic field at the source. Instead, it simply indicates the local wave propagation angle. This frequency limit, already used in paper I and section \ref{loc}, appears as a relevant upper limit for local sources. Whatever the frequency, the SPV-derived $\theta'(f)$ remains constant around $\sim75^\circ\pm15^\circ$.

In parallel with this method, based on the knowledge of the wave direction, an alternate technique based on polarization measurements only, and detailed in section \ref{polar}, was used to measure reliable $\theta'$ for frequencies around $f_\mathit{ce}$. We obtained $\theta'\ge78^\circ$ for local sources, that confirms and improves above results.

\subsection{Distant sources}

The SPV-derived beaming pattern of distant sources is investigated in Figure \ref{fig5}b, along the extended time-frequency interval [0700,1100]~UT and [7,1000]~kHz, where emission is highly circularly polarized (Figure \ref{fig1}b). 

LH (black symbols, V~$\ge0$) and RH (blue symbols, V~$\le0$) polarized emissions were previously identified as X and O modes on the basis of the comparison of their intensity between 15 and 30~kHz. At these frequencies, X and O mode components display similar $\theta'$ and comparable median values (solid lines). However, close to 10~kHz, some RH events display intensities comparable to LH ones. As their estimated beaming angle is close to $90^\circ$ (blue diamonds), they can be attributed to X mode, whose degree of circular polarization changes sign at $\theta'=90^\circ$. At frequencies higher than 40~kHz, where X mode prevails, the sign of V does not reverse at $90^\circ$, which suggests that beaming angles higher 90$^\circ$ are not physical here and simply illustrate the statistical noise. 

We now focus on the X mode beaming angles. Extending results of Figure \ref{fig5}a, the median (gray solid) remains approximately constant at $\theta'(f)=75^\circ\pm15^\circ$ up to 300~kHz and then decreases with frequency toward $\theta'(1000$~kHz)~$=50^\circ\pm25^\circ$. In appendix \ref{simu}, we quantify with simulations the observational bias induced by the employed radio localization technique. In particular, we show that the median displayed here (computed over all LH measurements) gives a reliable estimate of real beaming angles, while the large scattering, increasing with frequency, can be attributed to the instrumental noise on {\bf k} directions and the observational geometry.

In addition to the above direct determination, we also used an indirect technique to infer $\theta'$ for distant sources, that consists of modeling the narrowband intense part of the SKR low frequency envelope of Figure \ref{fig1}a, based on the PRES simulation code (Planetary Radio Emissions Simulator) \citep{Hess_GRL_08,Lamy_JGR_08b}. Here, we postulated SKR sources at frequencies covering the entire observed frequency range, along field lines located between 0030 and 0130~LT with invariant latitudes of $-80^\circ$. As PRES assumes straight line propagation, $\theta'$ was then used a (sensitive) free parameter to fit the low frequency envelope of the SKR spectrum between 10 and 60~kHz during the whole interval, as viewed mostly out of the sources. Best fit of the emission was obtained for $\theta'\sim75^\circ$, with an angular width of the cone of a few degrees. This confirms SPV-derived $\theta'$ between 10 and 60~kHz.

These results can be compared to previous direct or indirect measurements of SKR beaming angles. Below 500~kHz, the median behavior of Figure \ref{fig5}b is marginally consistent with the $50^\circ\pm10^\circ$ range derived for southern SKR sources by \citet{Cecconi_JGR_09} from 12~h of close mid-latitudes observations. This could result from $\theta'$ variable with time, position and/or azimuth with respect to the magnetic field, as well as affected by refraction along the ray path, expected to play a non negligible role at low frequencies. A statistical study is requested to answer these questions. Interestingly, Figure \ref{fig5}b is in excellent agreement with the oblique frequency-decreasing beaming angle computed by \citet{Lamy_JGR_08b} to model features in SKR dynamic spectra observed from equatorial latitudes at various distances. 

% As present measurements were acquired from very high latitudes, the tenuous medium along the ray path may induce less refraction than for close mid-latitude observations. 

\subsection{Intrinsic and apparent beaming pattern}

At Earth, AKR has been observed to be amplified quasi-perpendicularly to the local magnetic field within its source region \citep{Hilgers_JGR_92}, from unstable trapped \citep{Louarn_JGR_90}, or shell-type \citep{Ergun_ApJ_00} electron distributions. Then, refraction along the path, at the boundaries of auroral cavities or through the terrestrial plasmasphere, results in a complex oblique cone of emission, non-axisymmetric and partially filled \citep{Louarn_PSS_96a,Louarn_PSS_96b,xiao_JGR_07,Mutel_GRL_08}.

Here, local sources below $f_\mathit{ce}+1$~kHz display large $\theta'$, measured between $30^\circ$ and $80^\circ$ from direction measurements, or above $78^\circ$ from polarization measurements, consistent with perpendicular emission within error bars. This result is supported by conclusions drawn in companion studies: in paper I, we inferred perpendicular emission from the comparison of SKR cutoff frequencies and CMI resonance frequencies. In parallel, \citet{Mutel_GRL_10} computed CMI growth rates using ring-type electron distributions fitted to shell-like structures observed within the source region \citep{Schippers_JGR_10}, that account for the observed SKR intensity by being maximal for perpendicular amplification. In addition, we can notice that $\theta'\ge70^\circ$ together with a cone width of $5^\circ$, as previously estimated from far observations \citep{Lamy_JGR_08b}, yields a solid angle of $\sim0.5$~sr, anticipated in section \ref{efficiency}.

%Moreover, this result is also reinforced by the similarity with the case of AKR perpendicular emission from ring or shell distributions.

Above $f_\mathit{ce}+1$~kHz, or equivalently for distances beyond $\sim0.5$~R$_\mathit{S}$, beaming angles remain constant at $\theta'=75^\circ\pm15^\circ$, considered as a reliable estimate, up to 300~kHz, then decreasing with frequency. \citet{Lamy_JGR_08b} modeled such a frequency-dependent beaming angle from a loss cone electron distribution. As CAPS measurements rather suggest shell or ring-type distributions, that favor perpendicular emission, alternate possibilities must be explored. Emitted radio waves are likely to undergo refraction within a few tenths of kronian radii until they eventually reach $\theta'\sim75^\circ$ in regions where $N\sim1$. Indeed, in Figure \ref{fig3}, the locus of local sources precisely matches the footprints of magnetic field lines that were crossed by Cassini while assuming straight line propagation. This confirms that refraction must occur very close to the amplification region. We can make simple estimates of refractive conditions along the observed ray paths: assuming $\theta'=75^\circ$ with $N=1$ for distant sources, the Snell-Descartes law requires $N\sim0.966$ at the source for a strictly perpendicular emission.

Finally, the striking similarity of X and O mode beaming angles is intriguing. If both modes emanate from the same regions, as suggested by Figure \ref{fig3}, at frequencies close to $f_\mathit{ce}$, itself much larger than $f_\mathit{pe}$, refraction is expected to be more significant for the X mode close to $f_\mathit{X}$ ($N_\mathit{_X}<1$), than for the O mode far above $f_\mathit{O}$ ($N_\mathit{_O}\sim1$).

A dedicated ray tracing study is required to quantify the SKR ray path over all accessible directions for accessible magneto-ionic modes. Interestingly, we note that the jovian decametric emission (DAM) also displays beaming angles about $70^\circ$ \citep{Queinnec_JGR_98}.

%At Earth, AKR has been observed to be emitted quasi-perpendicularly to the magnetic field within its source region, consistent with CMI perpendicular generation from trapped, or shell-like unstable electron distributions. Measurements were performed with spinning spacecraft, by identifying the minimum of the time-variable amplitude of the wave electric field.
% The nature of the electron distribution responsible for SKR generation is crucial, since a loss-cone feature, as the one used in \citep{Lamy_JGR_08b} naturally leads to oblique emission, whereas ring or shell distributions, yield perpendicular emission, as the AKR at Earth.
%- Wavelength at 10~kHz=30000km (half a radius)

\begin{figure*}
\centering\includegraphics[width=0.95\textwidth]{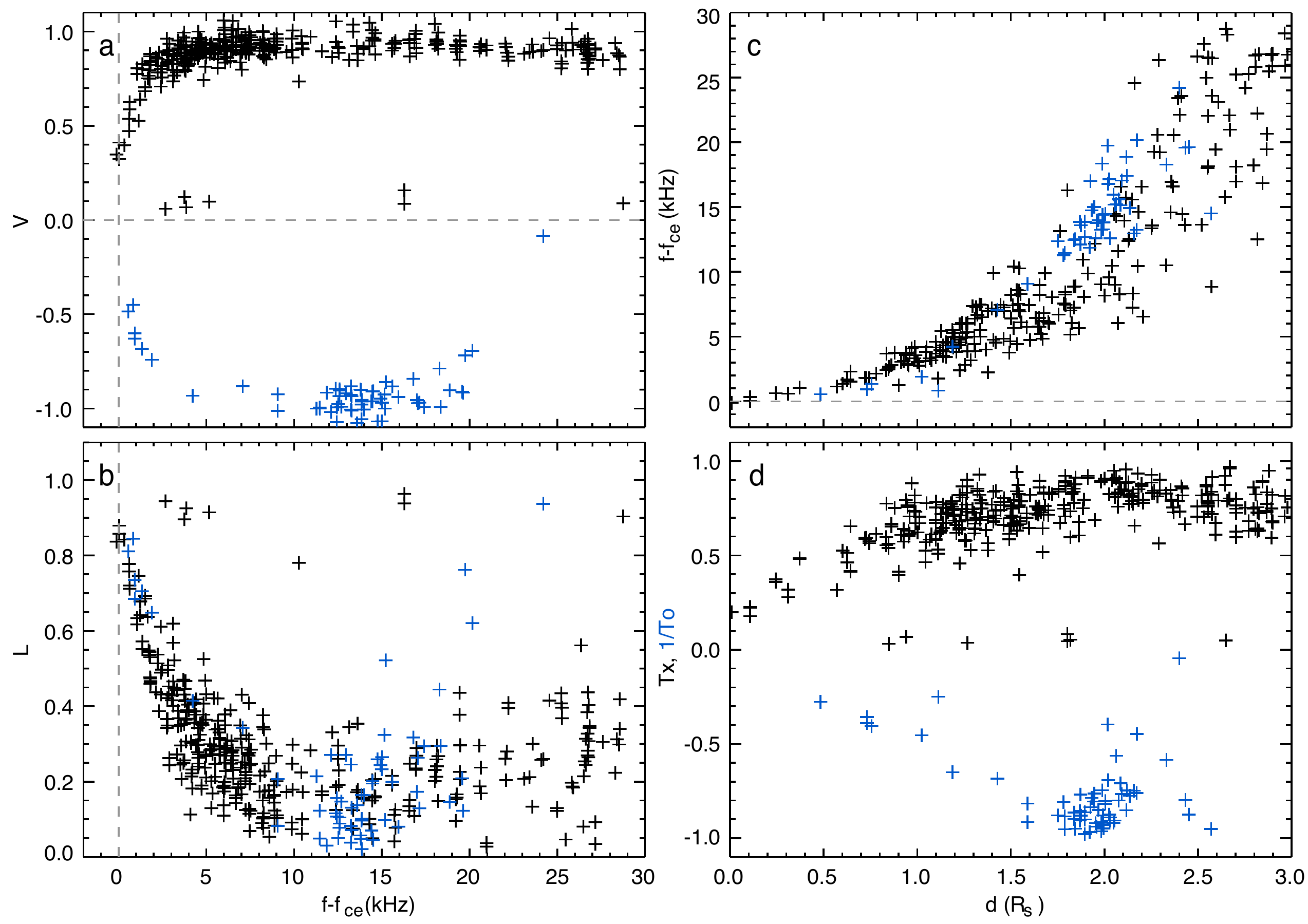}
\caption{Polarization state of radio waves observed between 0712 and 0912~UT (source region extended by one hour to include significant O mode emission) and $f_\mathit{ce}-2$~kHz~$\le f\le f_\mathit{ce}+30$~kHz. (a) Circular and (b) linear degree of polarization V and L~=~(Q$^2+$~U$^2)^{1/2}$ are displayed as a function of $f-f_\mathit{ce}$, whereas (c) $f-f_\mathit{ce}$ and (d) axial ratio of the polarization ellipse T~=~[(Q$^2$+U$^2$+V$^2$)$^{1/2}-$~L]/V \citep{Kraus_66} are displayed as a function of the distance to the source (the distance along which the wave direction intercepts the surface $f=f_\mathit{ce}$). Black and blue symbols refer to V~$\ge0$ and V~$\le0$ respectively, identified as X and O modes in section \ref{magnetomodes}. As these modes are expected to be orthogonal (T$_\mathit{_X}$T$_\mathit{_O}=-1$), Figure \ref{fig6}d superimposes T$_\mathit{_X}$ and 1/T$_\mathit{_O}$, so that polarization ellipses can be directly compared. RPWS data have again been selected with the selection described in section \ref{data}, with an additional criterion retaining only measurements with a normalized degree of total polarization (Q$^2$+U$^2$+V$^2$)$^{1/2}$ between 0.85 and 1.1. Moreover, considering that local sources correspond to $f\le f_\mathit{ce}+1$~kHz, as illustrated in Figures \ref{fig3} and \ref{fig5}, associated distances have been systematically set to 0, even when not crossing the surface $f=f_\mathit{ce}$. X and O modes follow a similar trend in all panels and $d$ regularly increases with frequency, as expected. SKR waves are observed to be quasi-purely circularly polarized above $f_\mathit{ce}+10$~kHz ($d\ge1.5$~R$_\mathit{S}$), and elliptical below, toward V~$\sim0.2-0.4$, L~$\sim0.8-0.9$ and T$_\mathit{_X}$~$\sim0.1-0.3$ at $f=f_\mathit{ce}$.}
\label{fig6}
\end{figure*}

\section{State of polarization}
\label{polar}

\subsection{SKR polarization ellipse along the propagation}
\label{observed_polar}

Previous studies showed that SKR is quasi-fully polarized, with a degree of total polarization (V$^2$~+~Q$^2$~+~U$^2)^{1/2}\sim1$, and that the observed polarization evolves with the location of the observer \citep[and references therein]{Fischer_JGR_09}. Whereas SKR polarization is quasi-purely circular when detected from low latitudes ($\le30^\circ$), it becomes mainly and significantly elliptical when observed from higher latitudes (between $30^\circ$ and $60^\circ$). 

Here, Cassini's trajectory swept very high southern latitudes, from $-40^\circ$ to $-75^\circ$, along which two important features are brought to light by Figure \ref{fig1}b. First, distant sources above $f_\mathit{ce}$ display high levels of circular polarization with $|$V$|\sim1$ (in regions of the spectrogram where GP results are not affected by the geometric configuration between antennas and real wave directions). This strongly differs from the trend noticed by \citet{Fischer_JGR_09} in the range $-40^\circ$ to $-60^\circ$. Second, as noticed in section \ref{magnetomodes}, circular polarization vanishes for local sources, close to $f_\mathit{ce}$, whereas we checked that radio waves remain quasi-fully polarized. This suggests that the wave polarization changes along the ray path with the distance of propagation.

Figure \ref{fig6} investigates the SKR full state of polarization in detail. As mentioned earlier, 3-antenna measurements bring a direct tridimensional determination of the polarization ellipse. On the left-hand side, the degrees of circular (V, Figure \ref{fig6}a) and linear (L, Figure \ref{fig6}b) polarization are plotted as a function of $f-f_\mathit{ce}$. On the right-hand side, $f-f_\mathit{ce}$ (Figure \ref{fig6}c) and the axial ratio of the polarization ellipse (T, Figure \ref{fig6}d), are plotted as a function of the distance to the source $d$. Figure \ref{fig6}c illustrates that the quantity $f-f_\mathit{ce}$ is a proxy for $d$. The dataset employed here retains only quasi-fully polarized events with (V$^2$~+~Q$^2$~+~U$^2)^{1/2}\ge0.85$, observed between 0712 and 0912~UT (to include significant X and O mode contributions), and frequencies below $f_\mathit{ce}+30$~kHz ($d$ between 0 and 3~R$_\mathit{S}$). 

SKR polarization continuously evolves with $f-f_\mathit{ce}$. Above 10~kHz, the observed radio sources are located beyond $1.5$~R$_\mathit{S}$ and are highly circularly polarized, with $|$V$|$~$\sim0.8-1.1$, L~$\sim0.0-0.3$ and T~$\sim0.7-1.0$. A few events, seen at all frequencies with low V , high L and low T, correspond to unreliable GP results (located in time-frequency regions of weak signal on Figure \ref{fig1}b). Below 10~kHz toward decreasing $f-f_\mathit{ce}$, a continuous trend shows more and more elliptically polarized emission, reaching V~$\sim0.2-0.4$, L~$\sim0.8-0.9$ and AR~$\sim0.1-0.3$ at the limit $f=f_\mathit{ce}$. Stokes parameters are very similar for X and O modes, whatever the frequency. This is not surprising since characteristic polarizations are expected to be orthogonal (T$_\mathit{_X}$T$_\mathit{_O}=-1$, see section \ref{waves}), which means that polarization ellipses of X and O mode only differ by a rotation of $\pi/2$, the former having its main axis along the magnetic field direction.

In summary, SKR polarization is naturally strongly elliptical at the source, before being circularized along its propagation path. These two aspects are investigated below.

\subsection{Intrinsic polarization}

\begin{figure}
\centering\includegraphics[width=0.45\textwidth]{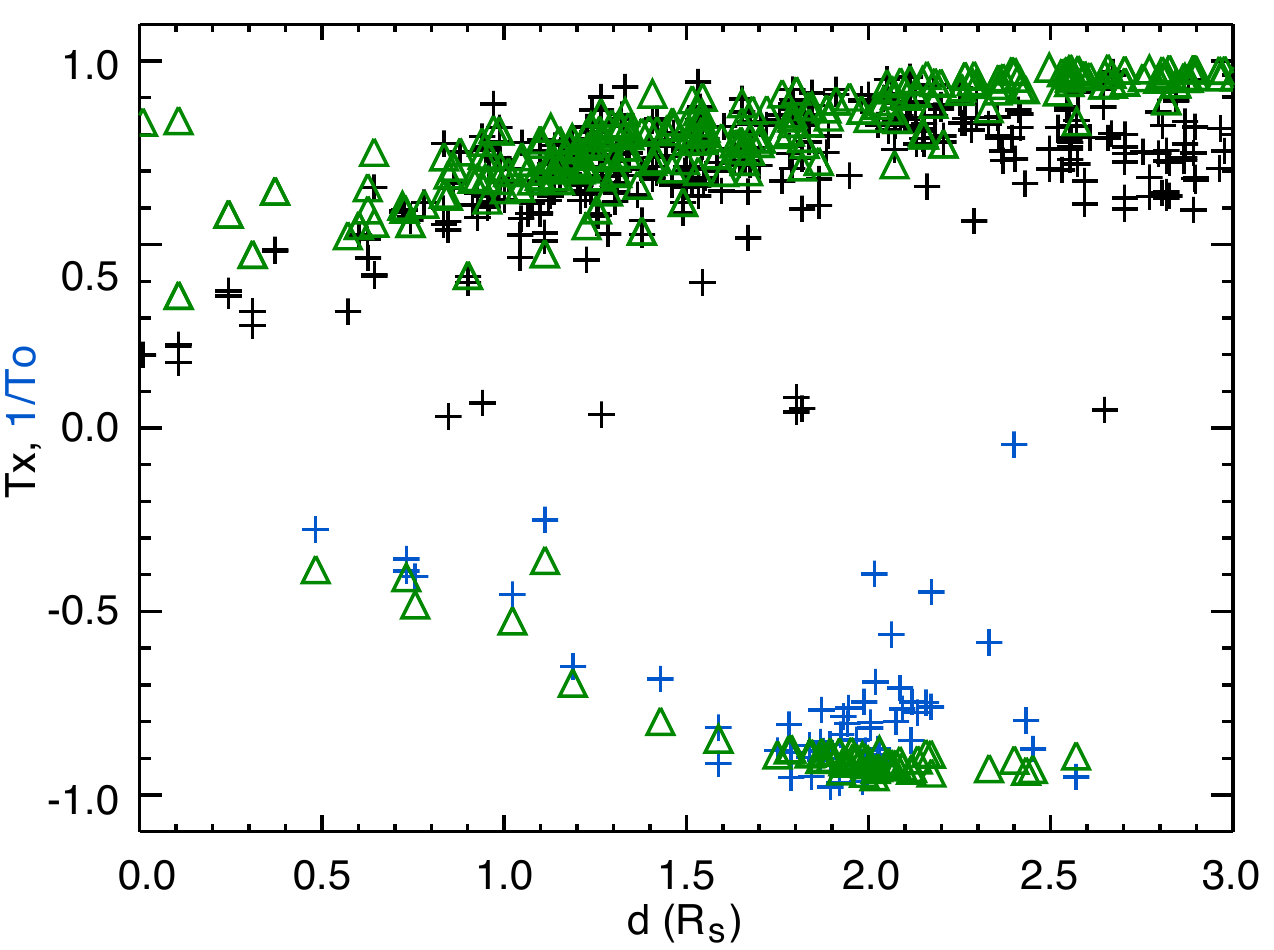}
\caption{Identical to Figure \ref{fig6}d, except that axial ratios predicted by equation \ref{eq_AR}, using $X$, $Y$ and local wave propagation angles $\theta'$, have been overplotted with green triangles. The overall correspondence between both distributions is clear.}
\label{fig7}
\end{figure}

In contrast to the common picture of planetary auroral radio emission being circularly polarized, SKR naturally displays elliptical polarization at the source. How does this compare to predictions of the magneto-ionic theory?

Using conditions prevailing within a CMI source ($X\ll 1$, $Y\sim1$ and weakly relativistic electrons), \citet{Melrose_AA_91} showed that the polarization of a cyclotron emission and the one of the X mode coincide and reduce to T~$\sim\cos{\theta'}$ to first order. Here, we can use the exact expression given by equation \ref{eq_AR} with local measurements of T$_\mathit{_X}$, X and Y to directly compute the beaming angle at the source. For closest sources in Figure \ref{fig6}d, T$_\mathit{_X}\le 0.2$ yield $\theta'\ge78^\circ$, supporting quasi-perpendicular emission suggested in section \ref{beaming}. 

The relation T$\sim\cos{\theta'}$ is not valid for distant sources, for which Y shifts from unity. As an illustration, while axial ratios increase with $d$ (Figure \ref{fig6}d), $\theta'$ remains constant (Figure \ref{fig5}). Nonetheless, the knowledge of $X$, $Y$ as well as the local wave propagation angle (green $\theta'$ in Figure \ref{fig5}) for each time-frequency measurement allows us to directly use equation \ref{eq_AR} to predict the local value of T$_\mathit{_{X}}(f)$ and 1/T$_\mathit{_{O}}(f)$ for all waves (including local and distant sources), whatever their ray path.

Observed and predicted axial ratios are superimposed in Figure \ref{fig7}. Though some waves propagate through distances of a few kronian radii, the overall correspondence is pretty good. Precisely, T${_\mathit{_X}}(f)$ and 1/T${_\mathit{_O}}(f)$ evolve similarly between 0.5 and 2~R$_\mathit{S}$. Below 0.5~R$_\mathit{S}$, predicted T${_\mathit{_X}}$ are slightly higher than observed ones, which is the consequence of underestimated MAG-derived $\theta'(f)$ for local sources ($\sim70^\circ$). Above 2.5~R$_\mathit{ce}$, the discrepancy between observed and predicted T$_\mathit{_X}$ corresponds to noisy results observed close to 40~kHz in Figure \ref{fig1}b. In summary, the magneto-ionic theory satisfactorily accounts for the observed SKR elliptical polarization.

\begin{figure}
\centering\includegraphics[width=0.45\textwidth]{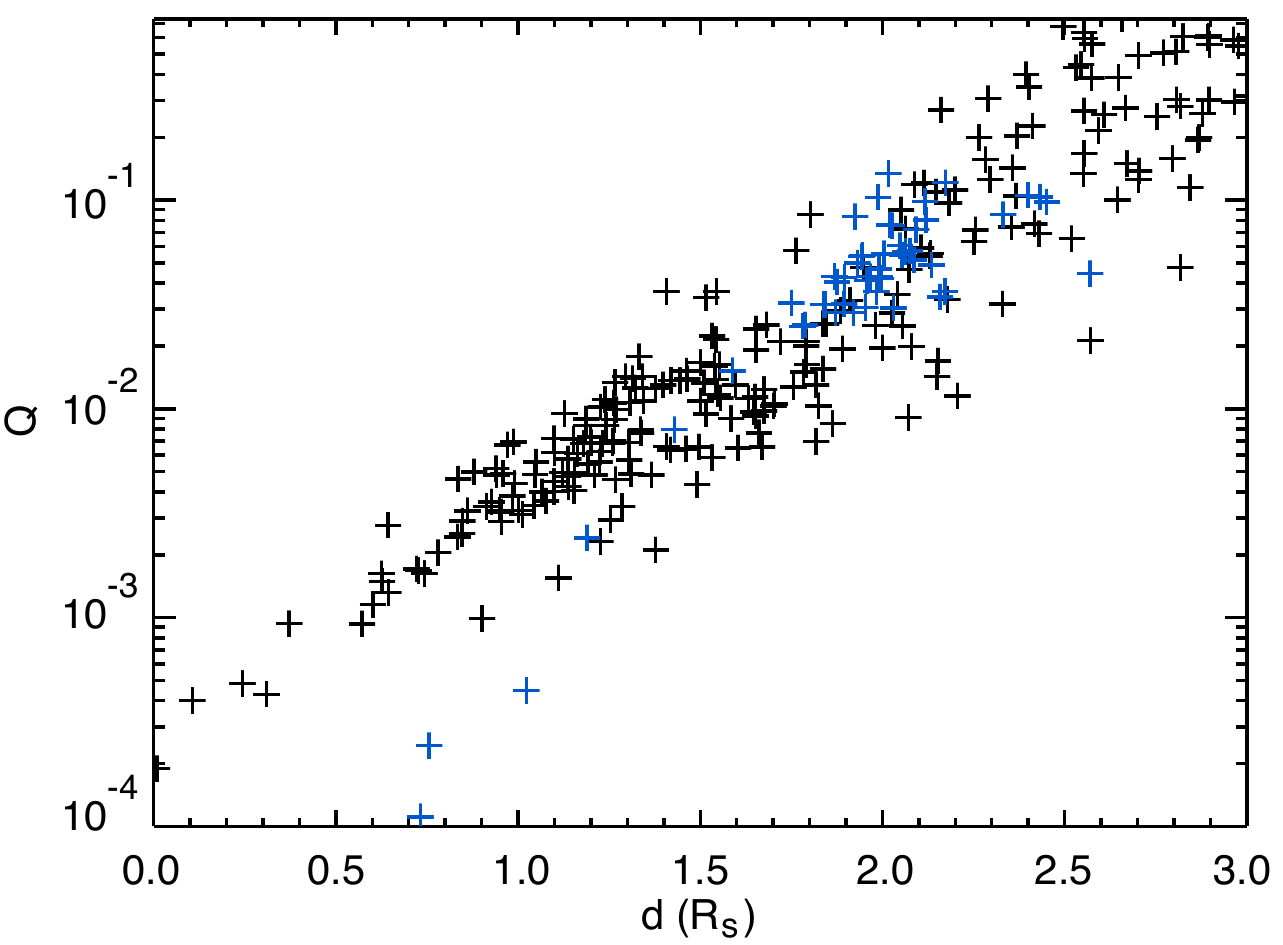}
\caption{Coupling factor Q \citep{Sawyer_JGR_91}, computed from equation \ref{coupling_factor} and data of Figure \ref{fig6}d, as a function of $d$. Q remains small whatever the distance of propagation.}
\label{fig8}
\end{figure}

\subsection{Polarization transfer}

The above result can be analyzed in the frame of the mode coupling theory \citep{Booker_RS_36,Budden_RS_52,Cohen_ApJ_60,Melrose_80,Daigne_AA_84}, used to investigate the elliptical polarization of jovian decametric emission (DAM) \citep{Goertz_PSS_74,Lecacheux_AA_91,Melrose_AA_91,Shaposhnikov_AA_97} and proposed to analyze first results on SKR elliptical polarization \citep{Fischer_JGR_09}.

When propagating through an inhomogeneous plasma, the polarization transfer of a radio wave can be described by two limiting regimes. If the medium is changing slowly enough ($X$, $Y$ and $\theta'$ do not change significantly over a wavelength), it can be considered as locally homogeneous and characteristic magneto-ionic modes are ''weakly coupled''. This means that natural modes are preserved and propagate independently of each other. The magneto-ionic theory prevails and their polarization and index of refraction, described by equations \ref{eq_N}-\ref{eq_AR}, slowly evolve along the ray path. For instance, their sense of circular polarization can reverse when {\bf k.B} changes sign ($\theta=\pi/2$ in equation \ref{eq_AR}), and Faraday rotation can occur. In contrast, if the medium changes rapidly, characteristic modes become "coupled" and the magneto-ionic theory does not apply any more. Characteristic modes do not propagate independently and can transfer energy to each other, so that the polarization of a given wave is retained along the ray path. The radiation propagates as in vacuum. 

The transition between regions of weak and strong coupling has been interpreted as a limiting polarization zone (LPZ), beyond which the wave polarization is fixed \citep{Budden_RS_52}. It has been characterized by \citet{Cohen_ApJ_60} in the high frequency approximation ($X,Y\ll 1$) in terms of a transitional frequency $f_t$ whose expression, established in the cases of quasi-longitudinal (QL) and quasi-transverse (QT) propagation, is proportional to the plasma density and the magnetic field strength (see precise expressions in equations 53 and 58). \citet{Cohen_ApJ_60} showed that weak coupling occur for $f\ll f_t$ or $f^4\ll f_t^4$ respectively for QL or QT propagation, and strong coupling when the opposite condition is fulfilled. For the jovian magnetosphere, \citet{Lecacheux_PRE_88} reduced $f_t$ to a function of the electron density $n$, and showed that strong coupling at 30~MHz requires $n\le5$~cm$^{-3}$. A similar result was obtained by \citet{Melrose_AA_91} for a region close to the source, using $Y\sim1/2$ and a typical length of inhomogeneity of one jovian radius, and then adapted to Saturn by \citet{Fischer_JGR_09} to infer $n\le0.01$~cm$^{-3}$ for SKR frequencies below 1~MHz. Transposed to the equivalent frequency of 20~kHz in the present study ($Y\sim1/2$ and $f_\mathit{ce}\sim10$~kHz), the region of strong coupling shall be reached for $n\le0.001$~cm$^{-3}$, a much lower value than the 0.01~cm$^{-3}$ measured locally (see paper I). This qualitatively shows that the LPZ at 20~kHz has not been reached yet and that associated radio waves propagate in a region of weak coupling. In addition, it is unlikely to find such low electron densities within the kronian magnetosphere, so that the LPZ may not be reached until waves eventually go out of the magnetosphere.

To estimate precisely the amount of coupling along the ray path, it is more convenient to use the coupling factor $Q$ introduced by \citet{Budden_RS_52}, and defined as the ratio of the rate at which the polarization of one characteristic mode changes along the ray path to the rate at which characteristic modes get out of phase. For instance, \citet{Sawyer_JGR_91} explicited $Q$ as:

\begin{equation}
Q=\frac{1}{2}\frac{\partial T/\partial d(T^2-1)^{-1}}{2\pi f|\Delta N_\mathit{_{X,O}}|/c}
\label{coupling_factor}
\end{equation}

where $\Delta N_\mathit{_{X,O}}=|N_\mathit{_O}-N_\mathit{_X}|$. Regions of weak and strong coupling then correspond to $Q\ll 1$ and $Q\gg 1$, respectively. Here, $\partial$T/$\partial$d can be simply estimated from Figure \ref{fig7}a, where T$_\mathit{_{X}}$ and 1/T$_\mathit{_{O}}$ vary approximately linearly by an amount of 0.7 over a distance of 2~R$_\mathit{S}$, while $\Delta N_\mathit{_{X,O}}$ is computed from equation \ref{eq_N} from $X, Y$ and $\theta'$.

Figure \ref{fig8} shows $Q(d)$ computed for X and O mode emissions displayed in Figures \ref{fig6}d and \ref{fig7}. This result was checked by using alternate coupling factors derived in other polarization studies \citep{Daigne_AA_84,Melrose_AA_91,Shaposhnikov_AA_97}, that all give similar or lower values of Q. Low coupling factors are thus observed at all frequencies, with $d\le3$~R$_\mathit{S}$, and consequently characterize a region where characteristic modes are not or weakly coupled. SKR is mainly emitted and remains in the X mode, with characteristics described by the magneto-ionic theory. An important consequence is that Faraday rotation is unlikely to occur along the investigated ray paths.

\subsection{Consequences and expectations for other planetary auroral radio emissions}

Among planetary auroral radio emissions, elliptical polarization was unambiguously observed only for jovian DAM and kronian SKR, whereas AKR at Earth or UKR at Uranus where characterized as being mostly circular, as reminded by \citet{Fischer_JGR_09}. 

At Jupiter, early DAM studies proposed that the observed elliptical polarization could be fixed by strong coupling at the LPZ in the close neighborhood of the source \citep{Lecacheux_AA_91,Melrose_AA_91}, in which case the wave polarization would carry information on the source region and the generation mechanism. Alternately, \citet{Shaposhnikov_AA_97} proposed that elliptical polarization might be formed by moderate linear mode coupling in transitional regions (TR) outside of the source region, defined by $A^2\sim1$ at the transition between QT and QL propagation, and carries little information on the source region. Here, $X=0.01$, $\theta'=75^\circ$ and $Y$ varying from 1 to 1/2 corresponds to $A^2$ between 3.3 and 0.8. Thus, when propagating out of their source region, SKR waves precisely lies within the transitional region where weak mode coupling occurs.

At Earth, quantitative measurements of AKR polarization with several spacecraft have established that the emission is strongly circularly polarized, whatever the location of the observer \citep{Lefeuvre_AG_98,Panchenko_RS_08,Lamy_JGR_10}. At the light of SKR polarization properties, the non detection of elliptical AKR is intriguing. From the expression of the transitional frequency transposed to Earth, weak coupling is expected for $n\ge0.1$~cm$^{-3}$ at the AKR frequency peak of 200~kHz, a condition which is fulfilled both within ($n\sim1$~cm$^{-3}$) and outside ($n\sim5$ to 10~cm$^{-3}$) the AKR source region \citep{Roux_JGR_93,Ergun_GRL_98}. We consequently expect AKR polarization to vary as the one of the characteristic mode on which it is emitted. Owing to similar conditions of generation for SKR and AKR (large $\theta$, $Y\sim1$ and $X\ll 1$), the AKR polarization at the source should be strongly elliptical at least within auroral cavities. 

%The non-detection of AKR elliptical polarization may simply result from the absence of quantitative polarization measurements within auroral cavities. Out of the latter, the sudden increase in density for an oblique wave propagation angles shall lead to axial ratios close to 1.

At Uranus, a polarization model based on the magneto-ionic theory was developed to derive the limiting polarization of Uranian Kilometric Radiation (UKR) \citep{Sawyer_JGR_91}, and successfully retrieved axial ratios derived from a few 2-antenna measurements of Voyager. This strengthens the picture of planetary radio emissions polarized accordingly to the conditions encountered along the ray path.

Coming back to Saturn, considering that SKR emission is beamed in a thin cone of emission aligned with the local magnetic vector, the plasma parameters $X$, $Y$ and the wave propagation angle $\theta'$ will change differently with the azimuthal angle around {\bf B} (in spherical coordinates), as well as the source locus. For instance, a radio wave propagating toward the equatorial dense plasma will encounter increasing densities, preventing strong coupling and circularizing the wave polarization, in agreement with near-equatorial observations below 30$^\circ$ latitude. As the index of refraction and the wave polarization are expected to vary significantly with azimuth, invariant latitude and local time, a detailed modeling of the SKR polarization transfer in appropriate plasma conditions, beyond the scope of this paper, is required to determine why elliptical polarization is well detected over the whole SKR spectrum for mid-latitudes observations and which information on the source is carried by observed waves.

\begin{figure*}
\centering\includegraphics[width=0.8\textwidth]{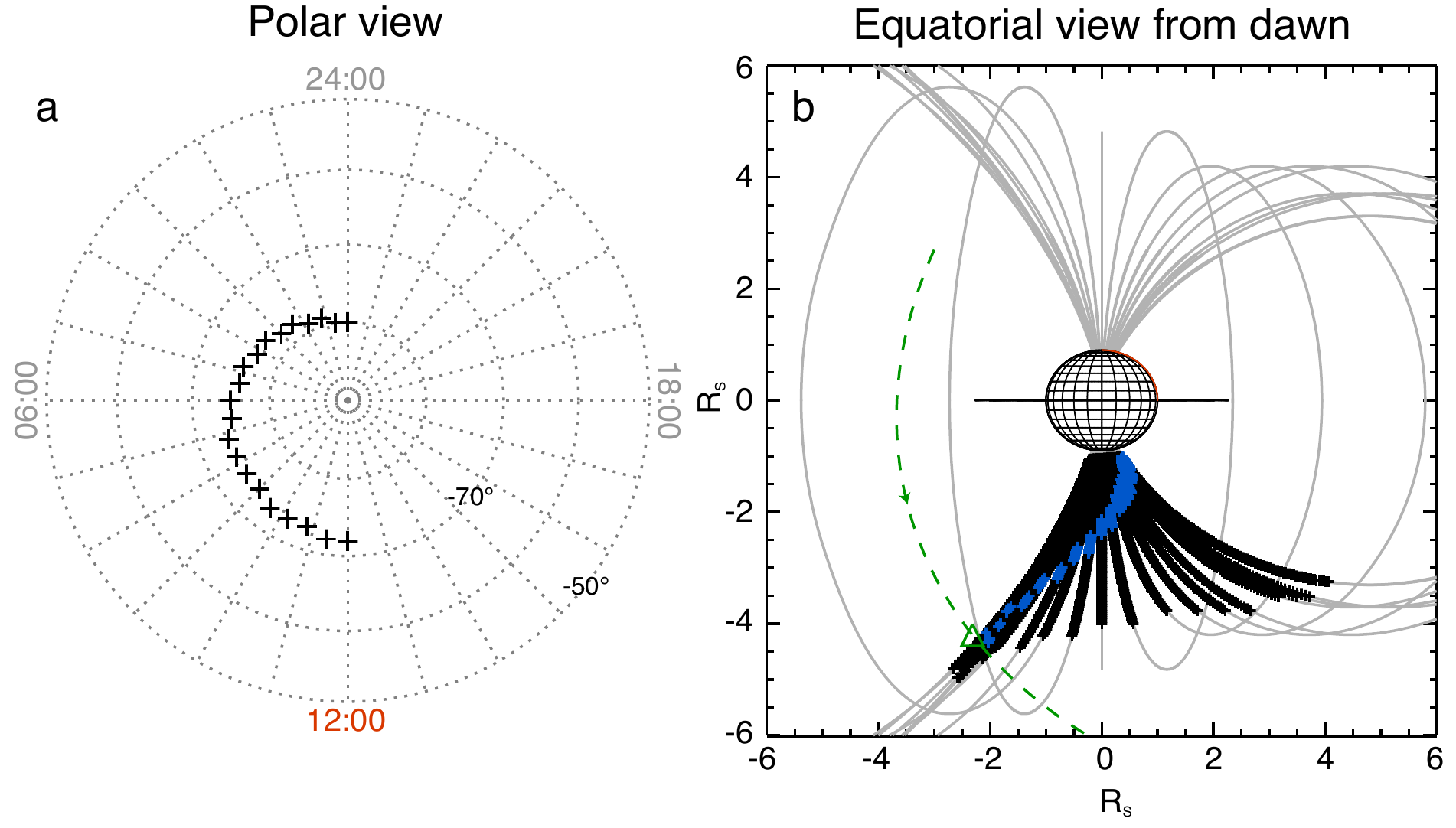}
\caption{(a) Southern polar view of a simple spiral auroral oval, comparable to the one of Figure \ref{fig3}b. (b) Equatorial view from dawn of corresponding magnetic field lines (gray), with simulated radio sources distributed between 6 and 1000~kHz (black crosses). The location of the observer is set at $08$:$12$ (green triangle) along the Cassini's trajectory (green dashed). Blue crosses display radio sources visible from the spacecraft for a beaming angle $\theta'_0$ (here equal to $75^\circ\pm5^\circ$).}
\label{fig9}
\end{figure*}

\section{Summary}
\label{summary}

In this study, we measured the properties of SKR along its propagation path from its source region to the observer thanks to simultaneous radio (electric and magnetic wave components), magnetic field and electron measurements acquired by the Cassini spacecraft.

SKR is mainly emitted in the fast X mode, at frequencies above the X mode cutoff derived for the observed hot auroral plasma, dominated by 6 to 9~keV electrons. From in situ observations, we estimated the mean electron-to-wave energy conversion efficiency about 1\% (2\% peak), in agreement with previous macroscopic estimations. SKR is also observed in the O mode, typically 2 orders of magnitude fainter, whereas no kilometric Z mode was detected. 

X and O SKR components are found to emanate from the same regions, and display similar beaming pattern and polarization at comparable frequencies. The SKR beaming pattern has been estimated by various techniques. The radio localization technique allows us to measure a rough cone of emission with large aperture angle $\theta'=70^\circ\pm15^\circ$ below $f_\mathit{ce}+1$~kHz, corresponding to the upper frequency limit for local sources. But the most accurate beaming angles are given by direct polarization measurements yielding $\theta'\ge78^\circ$, which supports a quasi-perpendicular emission, consistently with results inferred in companion studies. For distant sources above $f_\mathit{ce}+1$~kHz, $\theta'(f)=75^\circ\pm15^\circ$ up to 300~kHz and decreases above to reach $\theta'(1000$~kHz)~$=50^\circ\pm25^\circ$. The rapid transition from quasi-perpendicular to oblique wave propagation angles could result from refraction along the ray path close to the source region. These results significantly differ from previous measurements of SKR beaming angles, likely to result from beaming angles varying with the location of the observer.

SKR polarization is shown to evolve with propagation, from strongly elliptical at the source to highly circular after propagation across 2~R$_\mathit{S}$ of high-latitude auroral plasma. Thanks to the knowledge of the local plasma parameters and wave propagation angle, we demonstrated that the polarization predicted by the magneto-ionic theory satisfactorily describes the observations. Conditions of weak mode coupling apply to the present event from the sources to a distance larger than 3~R$_\mathit{S}$, so that the SKR polarization corresponds to the one of characteristic modes. In addition, similarly to SKR, we predict that AKR polarization measured (at least) inside its source region shall be strongly elliptical.

Considering the numerous striking similarities between AKR and SKR, measured remotely and locally, the CMI process, responsible for AKR generation is reinforced as a universal mechanism responsible of planetary auroral radio emissions. The differences noticed between AKR and SKR source regions (frequency of emission, presence of auroral cavities, acceleration processes, observed polarization) are characteristic of the CMI operation in each magnetosphere.

% (see review of the possible mechanisms in \citep{Lequeau_JGR_84})

\appendix

\section{Simulation of radio source beaming pattern and observational bias}
\label{simu}

The determination of beaming angles from the radio localization technique used in this article has been described and discussed by \citet{Cecconi_JGR_09}. For radio waves observed at a frequency $f$, reliable $\theta'$ are obtained from directions that intercept the surface $f=f_\mathit{ce}$ (crosses in Figure \ref{fig5}), whereas only approximate $\theta'$ (generally close to $90^\circ$) can be derived from directions that do not intercept this surface (diamonds in Figure \ref{fig5}). These two distributions are generally distinct, as a result of the geometrical configuration between visible sources and the observer.

Here, we want to determine how Figure \ref{fig5}b is affected by this observational bias, and check if the median computed over all beaming angle measurements yields a reliable estimation of the real beaming angle.

To do so, we build a simple spiral auroral oval (Figure \ref{fig9}a), comparable to the one of Figure \ref{fig3}b. Along magnetic field lines mapping to this oval, we simulate southern radio sources distributed between 6 and 1000~kHz (Figure \ref{fig9}b). Then, we set the spacecraft location at $08$:$12$ (green triangle), corresponding to the entrance into the SKR source region, and identify which radio sources are visible for a fixed beaming angle $\theta'_0$ (blue crosses, here using $\theta'_0=75^\circ\pm5^\circ$) with a straight line propagation ($N=1$).

\begin{figure*}
\centering\includegraphics[width=0.8\textwidth]{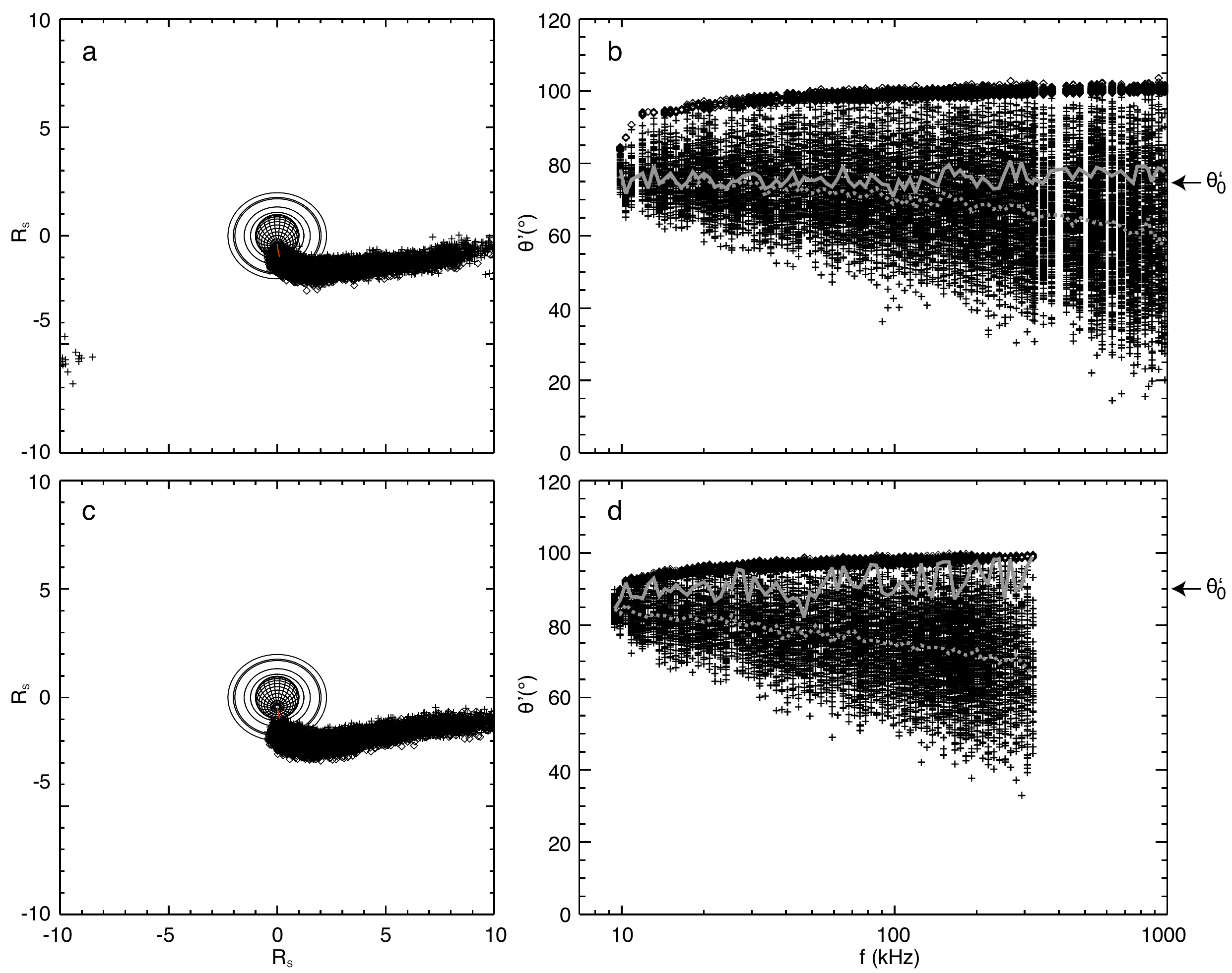}
\caption{(a) Blurred distribution of radio sources as viewed from the spacecraft with $\theta'_0=75^\circ\pm5^\circ$, comparable to Figure \ref{fig3}a. (b) Corresponding distribution of beaming angles, once processed by the radio localization technique. Similarly to Figure \ref{fig5}, crosses (diamonds) refer to directions of arrival that cross (do not cross) the surface $f=f_\mathit{ce}$. The dashed gray line corresponds to the median computed over reliable beaming angles (crosses) only, while the solid gray line gives the median computed over all beaming angles (crosses and diamonds). The latter is approximately equal to $\theta'_0$, whatever the frequency. (c,d) identical to (a,b) with $\theta'_0=90^\circ\pm5^\circ$.}
\label{fig10}
\end{figure*}

Afterwards, we model errors on visible sources by adding a gaussian noise distribution, made of 100 points and an angular width at half maximum equal to $2^\circ$ \citep{Cecconi_RS_05}, to each direction of arrival (Figure \ref{fig10}a,c). This blurred dataset is processed by the radio localization technique, and finally provides simulated beaming angles that directly compare to $\theta'_0$ (Figure \ref{fig10}b,d).

Figures \ref{fig10}a,b and \ref{fig10}c,d display results obtained for $\theta'_0=75^\circ\pm5^\circ$ and $\theta'_0=90^\circ\pm5^\circ$, respectively. As expected, visible radio sources lie on the dawnside, with a blurred distribution comparable to the one of Figure \ref{fig3}a, whereas slight differences of visibility are related to the value of $\theta'_0$. Our purpose here is not to strictly reproduce observations but to simply approach a realistic situation. 

The distributions of simulated beaming angles allow us to quantify the geometrical bias for observing conditions of Figure \ref{fig5}. First, the distribution of reliable beaming angles (crosses) naturally enlarges with frequency as a result of the distance to the source, with a median (gray dotted) decreasing with frequency. The latter artificially shifts from $\theta'_0$, which is only reached only at the lowest frequencies. Second, the distribution of approximate beaming angles (diamonds) is close to $90^\circ$ and slightly increases with frequency. Both of these features are well observed on Figure \ref{fig5}b.

Then, we note that the median computed over all beaming angles (solid gray) remains approximately constant and equal to $\theta'_0$ whatever the frequency and the value of $\theta'_0$, which means that it provides a reliable estimation of real beaming angles. As a result, we conclude that the median displayed in Figure \ref{fig5}b (approximately constant around $\sim75^\circ$ up to $\sim$300~kHz, decreasing at higher frequencies) cannot be explained by a geometrical bias and shows a real physical effect, either due to the driving of the CMI by a loss-cone electron distribution \citep{Hess_GRL_08,Lamy_JGR_08b} or to frequency-dependent refraction.

%\appendix{Relevance of the goniopolarimetric analysis}

%This justifies a posteriori the choice of a field model for remote radio localization technique \citep{Cecconi_JGR_09,Lamy_JGR_09}.

%Directions of arrival cannot be determined for purely linearly polarized waves.

%The assumption of transverse waves used in GP inversions is supported by two considerations. The longitudinal component is expected to be negligible in regions where $XY$ is low, including the case of SKR sources. Then the beaming angles derived from the polarizations at the source, are consistent with the ones given by GP inversion.

\begin{acknowledgements}
We wish to thank Cassini RPWS, MAG and CAPS engineers for support on instrumental questions. The French co-authors were supported by the CNES agency. The research at the University of Iowa is supported by NASA through Contract 1356500 with the Jet Propulsion Laboratory. We acknowledge the referees for accurate and useful discussions. 
\end{acknowledgements}

\end{article}

\end{document}